\documentclass[fleqn,usenatbib]{mnras}
\usepackage{newtxtext,newtxmath}
\usepackage[T1]{fontenc}

\DeclareRobustCommand{\VAN}[3]{#2}
\let\VANthebibliography\thebibliography
\def\thebibliography{\DeclareRobustCommand{\VAN}[3]{##3}\VANthebibliography}

\usepackage{graphicx}	
\usepackage{amsmath}	
\usepackage[dvipsnames]{xcolor}
\usepackage{enumitem}
\usepackage{url}
\usepackage{braket}
\usepackage{threeparttable}
\usepackage{hyperref}
\usepackage{multirow, tabularx}

\def\myerr[#1]{{\color{red} #1}}
\def\myemph[#1]{{\color{blue} #1}}
\def\hjmo[#1]{{\color{green} #1}}
\def\myrevise[#1]{{\bf #1}}
\def\revisestyle{}
\def\myrevises[#1]{{\color{red} #1}}
\def\term[#1]{{\bf \ttfamily #1}}
\def\B[#1]{{\bf #1}}

\setlist[itemize,1]{label=$\bullet$}
\setlist[itemize,2]{label=$\bullet$}
\setlist[itemize,3]{label=$\bullet$}
\setlist[itemize,4]{label=$\bullet$}
\setlist[itemize]{leftmargin=*}
\def\bit{\begin{itemize}[topsep=0em,parsep=0em,itemsep=0em,partopsep=0em,leftmargin=*]}
\def\bitt{\begin{itemize}[topsep=0em,parsep=0em,itemsep=0em,partopsep=0em,leftmargin=3.0em]}
\def\eit{\end{itemize}}
\def\benum{\begin{enumerate}[leftmargin=1em,itemsep=0pt,parsep=0pt,topsep=0pt]}
\def\eenum{\end{enumerate}}

\def\beq{\begin{equation}}
\def\eeq{\end{equation}}
\def\bey{\begin{eqnarray}}
\def\eey{\end{eqnarray}}

\def\bfrm[#1]{\mathrm{{\bf#1}}}

\def\gs{\mathrel{\raise1.16pt\hbox{$>$}\kern-7.0pt
		\lower3.06pt\hbox{{$\scriptstyle \sim$}}}}
\def\ls{\mathrel{\raise1.16pt\hbox{$<$}\kern-7.0pt
		\lower3.06pt\hbox{{$\scriptstyle \sim$}}}}
\def\smalldash{ \scalebox{0.5}[1.0]{\( - \)} }

\def\gtsima{\, {\buildrel > \over \sim} \,}
\def\ltsima{\, {\buildrel < \over \sim} \,}
\def\prosima{\, {\buildrel \propto \over \sim} \,}
\def\gsim{\lower.5ex\hbox{\gtsima}}
\def\lsim{\lower.5ex\hbox{\ltsima}}
\def\simgt{\lower.5ex\hbox{\gtsima}}
\def\simlt{\lower.5ex\hbox{\ltsima}}
\def\simpr{\lower.5ex\hbox{\prosima}}
\def\la{\lsim}
\def\ga{\gsim}

\def\specialname[#1]{\textbf{\textsc{#1}}}

\def\mpc{\, h^{-1}{\rm {Mpc}}}
\def\cmpc{\, h^{-1}{\rm {cMpc}}}

\def\gyr{\, h^{-1}{\rm Gyr}}

\def\gyri{\, h\,{\rm Gyr}^{-1}}

\def\msun{\, h^{-1}{\rm M_\odot}}

\def\Mhalo{M_{\rm h}}
\def\MhaloVec{ {\bf M}_{\rm h} }
\def\Rhalo{R_{\rm h}}
\def\Vhalo{ V_{\rm h} }
\def\Mstar{M_*}
\def\MstarInt{M_{\rm *,int}}
\def\MstarIntVec{{\bf M}_{\rm *,int}}
\def\zinfall{z_{\rm infall}}
\def\zanc{z_{\rm anc}}
\def\zrefer{z_{\rm ref}}
\def\zmerge{z_{\rm merge}}

\title[]{MAHGIC: A Model Adapter for the Halo-Galaxy Inter-Connection }

\author[Yangyao Chen et al.]{
	Yangyao Chen,$^{1,2}$\thanks{E-mail: yangyaochen.astro@foxmail.com}  
	H.J. Mo, $^{2}$   		
	Cheng Li, $^{1}$  		
	Kai Wang, $^{1,2}$		
	Huiyuan Wang, $^{3,4}$  
	Xiaohu Yang, $^{5,6}$   
	\newauthor
	Youcai Zhang $^{7}$ 	
	and Neal Katz $^{2}$    
	\\
	$^{1}$Department of Astronomy, Tsinghua University, Beijing 100084, China\\
	$^{2}$Department of Astronomy, University of Massachusetts, Amherst, MA 01003-9305, USA\\
	$^{3}$Key Laboratory for Research in Galaxies and Cosmology, Department of Astronomy, University of Science and Technology of China, Hefei, Anhui 230026, China \\
	$^{4}$School of Astronomy and Space Science, University of Science and Technology of China, Hefei, Anhui 230026, China\\
	$^{5}$Department of Astronomy, School of Physics and Astronomy, Shanghai Jiao Tong University, Shanghai, 200240, China\\
	$^{6}$Tsung-Dao Lee Institute, and Shanghai Key Laboratory for Particle Physics and Cosmology, Shanghai Jiao Tong University, Shanghai, 200240, China\\ 
	$^{7}$Key Laboratory for Research in Galaxies and Cosmology, Shanghai Astronomical Observatory, Shanghai 200030, China\\
}

\date{Accepted XXX. Received YYY; in original form ZZZ}

\pubyear{2020}

\begin{document}
\label{firstpage}
\pagerange{\pageref{firstpage}--\pageref{lastpage}}
\maketitle


\begin{abstract}
We develop a model to establish the interconnection between 
galaxies and their dark matter halos. We use Principal
Component Analysis (PCA) to reduce the dimensionality of both the mass assembly 
histories of halos/subhalos and the star formation histories of galaxies, 
and Gradient Boosted Decision Trees (GBDT)
to transform halo/subhalo properties into galaxy 
properties. We use two sets of hydrodynamic simulations to motivate 
our model architecture and to train the transformation. 
We then apply the two sets 
of trained models to dark matter only (DMO) 
simulations to show that the transformation is reliable and statistically
accurate. The model trained by a high-resolution hydrodynamic 
simulation, or by a set of such simulations implementing the same physics 
of galaxy formation, can thus be applied to large
DMO simulations to make `mock' copies of the 
hydrodynamic simulation. The model is both flexible 
and interpretable, {\revisestyle which paves the way for future applications 
in which we will constrain the model 
using observations at different redshifts simultaneously and explore 
how galaxies form and evolve in dark matter halos empirically}.
\end{abstract}

\section{Introduction}
\label{sec:intro}

In the framework of the $\Lambda$CDM cosmology, galaxies 
are assumed to form in dark matter halos produced by the gravitational 
instability of the cosmic density field
\citep{moGalaxyFormationEvolution2010a,wechslerConnectionGalaxiesTheir2018}. 
A key step in understanding 
the formation and evolution of galaxies is, 
therefore, to establish and understand the interconnection
between galaxies and dark matter halos. A variety of 
methods have been proposed and used to achieve this goal, 
such as: full numerical simulation that models subgrid physics numerically 
\citep[e.g.,][]{vogelsbergerPropertiesGalaxiesReproduced2014,
schayeEAGLEProjectSimulating2015,crainEAGLESimulationsGalaxy2015,
nelsonIllustrisTNGSimulationsPublic2019,
pillepichFirstResultsIllustrisTNG2018,springelFirstResultsIllustrisTNG2018,
nelsonFirstResultsIllustrisTNG2018,
naimanFirstResultsIllustrisTNG2018,marinacciFirstResultsIllustrisTNG2018},
matching galaxies and halos based on abundance 
\citep{moStructureClusteringLymanbreak1999,valeLinkingHaloMass2004,
guoHowGalaxiesPopulate2010,simhaTestingSubhaloAbundance2012}, 
clustering \citep{guoModellingGalaxyClustering2016}, and age 
\citep{hearinDarkSideGalaxy2013,
hearinDarkSideGalaxy2014,mengMeasuringGalaxyAbundance2020}, 
halo occupation distribution \citep{jingSpatialCorrelationFunction1998,
berlindHaloOccupationDistribution2002}, 
the conditional luminosity function \citep{yangConstrainingGalaxyFormation2003} and
conditional color-magnitude distribution \citep{xuConditionalColourMagnitude2018}, 
and empirical models based on the star formation histories of galaxies 
\citep{mutchSimplestModelGalaxy2013,luEmpiricalModelStar2014,luStarFormationStellar2015, 
mosterEmergeEmpiricalModel2018,behrooziUniverseMachineCorrelationGalaxy2019a,
mosterGalaxyNetConnectingGalaxies2020}.
These methods have yielded important results about the  
halo-galaxy interconnection. However, some issues still remain in such modeling.  
First, to build a successful model requires a systematic selection of 
significant halo features as predictors of the galaxy properties. 
The features selected should be such that they can be modeled reliably, 
are adaptable to new observational data, are non-redundant and yet sufficient 
to describe the data. Second, to simplify the mapping from halos to galaxies
needs some pre-processing of the input features to make the model easily 
interpretable. Finally, to suppress over-fitting of the data the model 
should be able to capture potential non-linearities in the halo-galaxy 
interconnection and should include an automated regularization. 
The deep neural network \citep[e.g.,][]{heDeepResidualLearning2016,huangDenselyConnectedConvolutional2017} 
provides a possible solution, but the structures of the networks 
need to be tuned and the physical interpretation underlying 
these structures is not straightforward.

{\revisestyle
Empirical models for the halo-galaxy relation in the 
literature can roughly be classified
into two categories according to their architecture.
Models in the first category, which may be called `transparent-box' models, 
include those based on abundance matching, 
clustering and age matching, conditional luminosity function, conditional 
color-magnitude distribution, and parametric mapping between halo and galaxy 
properties. These models are directly motivated 
by physical considerations, hence can be interpreted easily. 
However, these models usually adopt a set of halo features that may 
not be optimal for predicting galaxy properties owing to the loss of information 
and the degeneracy of halo properties. Furthermore, for models with 
parametric mapping, the number of features is usually limited because of 
the difficulty in creating multi-variate parametric formulations. 
This limits the flexibility of the model.
Models in the other category, which may be referred to as `black-box' models, 
usually use complex non-linear mappings, such as deep neural networks, 
to link halos and galaxies. These models can in principle be designed and 
tuned to make precise predictions for a large set of object and 
field properties, such as the baryon fraction in halos 
\citep{sullivanUsingArtificialNeural2018}, 
halo stellar and gas properties \citep{mosterGalaxyNetConnectingGalaxies2020, 
agarwalPaintingGalaxiesDark2018}, 
galaxy images \citep{zanisiRelationshipFineGalaxy2021}, 
and density fields \citep{horowitzHyPhyDeepGenerative2021, 
wadekarHInetGeneratingNeutral2020}.
Their major problem is that the complexity of the model  
can significantly hinder the interpretability. }

{\revisestyle
The gap between the `transparent-box' and `black-box' models 
motivates us to develop a scheme that takes advantages 
of both categories in their interpretability and flexibility. 
This can be achieved by using a deep neural network,
pruning its overly complicated parts, and replacing its mandatory 
parts with sub-models that have clearer physical interpretations. 
Since more pruning and replacing in general makes a model 
more transparent but less flexible, the pruning and replacing steps 
can be made adjustable, so that the model is an adapter
that allows a continuous transition between physically-motivated models 
and those based on modern deep learning.
}

{\revisestyle
To modify a deep neural network for our purpose, we can 
decompose the network into three parts along the information flow. 
The first, which can be viewed as a feature extractor,
is a set of layers that transforms the input of physical variables 
into a set of reduced hidden variables to simplify the regression problem.
The second, called the regressor, consists of set of layers
that maps the hidden input into a hidden output.  
The last part, which can be viewed as a feature restorer, is a set 
of layers that restores physical quantities from the hidden output of the 
regressor. Thus, the key to building our desired adapter 
is to prune and replace the three parts to make them interpretable.
}

In a recent paper, \citet[][thereafter, \textbf{Paper-I}]{chenHowEmpiricallyModel2021} 
developed an empirical method to link dark matter halos to central  
galaxies that form within them. They adopted a linear dimension reduction technique based on
Principal Component Analysis (\specialname[PCA]) and a tree-based model 
ensemble technique called Gradient Boosted Decision Trees 
(\specialname[GBDT]) to establish the halo-galaxy interconnection.
This method has the following advantages. First,  
by applying the \specialname[GBDT] regressors to simulated halos and galaxies, 
one can identify key halo properties that are the most relevant  
to the stellar properties of galaxies. This clearly 
demonstrates that the input features of an empirical model should be 
properly designed to avoid irrelevant and redundant halo properties.
Second, by using \specialname[PCA], the mass assembly history (MAH) of a dark matter halo, 
which in general is complex, can be described by a small number of principal 
components (PCs) without much information loss. 
As demonstrated in Paper-I and \cite{chenRelatingStructureDark2020},
PCs of subhalo MAH are not only tightly correlated with halo structural and environmental 
properties, but are also the most important quantities 
to predict the variance of  the star-forming main sequence. 
{\revisestyle
Similar techniques were used 
to reduce the dimensionality of halo MAH and 
find the principal direction in the halo property space 
\citep[e.g.][]{jeeson-danielCorrelationStructureDark2011a,wongWhatDarkMatter2012}, 
and to approximate the SFH of galaxies 
and find the principal direction in the galaxy color space 
\citep[e.g.][]{cohnCharacterizingSimulatedGalaxy2015,
cohnApproximationsGalaxyStar2018,
chaves-monteroSurrogateModellingBaryonic2020,
chaves-monteroSurrogateModellingBaryonic2021}. 
Finally, the use of \specialname[GBDT] regressors and classifiers 
can capture non-linear relations between halos and galaxies without
introducing significant over-fitting. 
The \specialname[GBDT] contains the non-linearity of the whole model 
in a single layer, making the model easily interpretable. 
All these advantages are important to take into account when 
modeling the halo-galaxy interconnection. In the model presented 
here, we therefore adopt \specialname[PCA] and \specialname[GBDT] 
for the pruning and replacing steps, respectively.
}

{\revisestyle The interpretability of the model} allows one to use both 
hydrodynamic simulations and observations to motivate and train the model. 
Depending on the training set, the model can be applied 
in two different ways. First, if the model is trained by a hydrodynamic simulation,
or by a set of simulations implementing the same set of physical processes, 
one can apply it to another, generally much larger, dark-matter-only (DMO) 
simulation to generate a copy using the dark matter halo population 
in the DMO simulation. This is an important application, because full 
hydrodynamic simulations are usually run with relatively small box size.
Such simulations may be sufficient to model galaxy formation in individual 
halos, but may not be able to provide a fair sample of the universe 
owing to the cosmic variance associated with the small simulation volume
\citep[see, e.g.,][]{somervilleCosmicVarianceGreat2004,mosterCOSMICVARIANCECOOKBOOK2011,
chenELUCIDVICosmic2019,mengMeasuringGalaxyAbundance2020}.
The copy made by the trained model combines the advantages 
of these two types of simulations, producing a much larger sample
statistically equivalent to the training set but 
with a much reduced cosmic variance.
Second, if observational data are used to constrain the model, the 
architecture learned from hydrodynamic simulations can be used for the model 
design, so that model parameters inferred from observations 
can be interpreted in terms of physical processes. 
For example, the mapping from halos to galaxies obtained from observational 
data can be used to reveal halo properties that are the most 
important in determining a given set of properties of the galaxy population.
The model inferred from the data can also be compared directly with 
that trained by the hydrodynamic simulation to test the assumptions
made in the simulation. {\revisestyle Another potential application, 
which is beyond the scope of this paper, is to use the constrained 
halo-galaxy relation (the constrained galaxy bias model) 
to populate large-box simulations to test precision cosmology.}

{\revisestyle
In this paper, we adopt the above adapter architecture and extend the model 
described in Paper-I for central galaxies to a full pipeline for the whole galaxy population. 
The pipeline, named \specialname[MAHGIC]-
a {\it Model Adapter for the Halo-Galaxy Inter-Connection}, 
is trained and tested here using hydrodynamic simulations, before we apply it to
observational data in the future. 
}
The paper is organized as follows. 
In \S\ref{sec:data} we introduce the numerical simulations, 
the halo properties, and the samples used in our analysis.
In \S\ref{sec:model}, we describe our model and how the 
different components of the model are pieced together into the pipeline
\specialname[MAHGIC]. We test the performance of the pipeline 
in \S\ref{sec:results} using hydrodynamic simulations. 
Our main results are summarized and discussed in \S\ref{sec:summary}.

\begin{center}
\begin{table*} 
\caption{Summary of the four simulations used in this paper. 
	Listed information includes 
	cosmological parameters, 
	box size $L_{\rm box}$, 
	number of resolution units $N_{\rm resolution}$, 
	dark matter particle mass $m_{\rm dark\ matter}$,
	and target baryon mass $m_{\rm baryon}$. 
	Cosmologies are taken from
	Planck15 \citep{adePlanck2015Results2016}, 
	WMAP5 \citep{dunkleyFIVEYEARWILKINSONMICROWAVE2009}, and 
	Planck13 \citep{adePlanck2013Results2014a}. 
	In TNG, $N_{\rm resolution}$ is the number of dark matter particles plus the 
	initial number of gas cells, and gas cells are refined or 
	de-refined such that their mass is kept within a factor of 
	2 of $m_{\rm baryon}$.
	In EAGLE, 
	$N_{\rm resolution}$ is the number of dark matter particles plus the 
	initial number of baryonic particles and $m_{\rm baryon}$ is the initial 
	baryonic particle mass.
	In TNG-Dark and ELUCID, $N_{\rm resolution}$ is the number of dark matter 
	particles.
	}
\begin{tabularx}{\textwidth}{c | X | >{\hsize=.15\hsize}X | >{\hsize=.15\hsize}X >{\hsize=.15\hsize}X >{\hsize=.13\hsize}X}
	\hline 		
		Simulation   			
		& Cosmology   
		& $L_{\rm box} $  \newline $[\cmpc]$
		& $N_{\rm resolution}$    
		& $m_{\rm dark\ matter}$   \newline $[\msun]$
		& $m_{\rm baryon}$			\newline $[\msun]$
		\\
	\hline\hline
		TNG 					
		& \multirow{3}{9.0cm}{
			Planck15: 
			$h=0.6774$, $\Omega_{\Lambda,0}=0.6911$, $\Omega_{M,0}=0.3089$, 
			$\Omega_{B,0}=0.0486$, $\Omega_{K,0}=0$, $\sigma_8=0.8159$, $n_s=0.9667$}
		& \multirow{3}{*}{75}
		& $2 \times 1820^3$
		& $5.1\times 10^6$ 
		& $9.4\times 10^5$
		\\ & & & & & \\
	\cline{1-1}\cline{4-6}
		TNG-Dark 	
		&
		& 
		& $1820^3$
		& $6.0\times 10^6$ 
		& -
		\\ & & & & & \\
	\hline
		ELUCID					
		& WMAP5: 
			$h=0.72$, $\Omega_{\Lambda,0}=0.742$, 
			$\Omega_{M,0}=0.258$, $\Omega_{B,0}=0.044$, $\Omega_{K,0}=0$, 
			$\sigma_8=0.80$, $n_s=0.96$
		& $500$
		& $3072^3$
		& $3.08 \times 10^8$
		& -	
		\\
	\hline
		EAGLE 
		& Planck13: 
			$h=0.6777$, $\Omega_{\Lambda,0}=0.693$, 
			$\Omega_{M,0}=0.307$, $\Omega_{B,0}=0.04825$, $\Omega_{K,0}=0$, 
			$\sigma_8=0.8288$, $n_s=0.9611$
		& $67.8$
		& $2 \times 1504^3$
		& $6.57 \times 10^6$
		& $1.23 \times 10^6$
		\\
	\hline
\end{tabularx}
\label{tab:simulations}
\end{table*}
\end{center}

\section{The Data}
\label{sec:data}

\subsection{The Simulations}
\label{ssec:simulations}

Throughout this paper, we use four simulations to motivate, 
build, train, and test our pipeline, \specialname[MAHGIC].

The first is Illustris-TNG 
\citep{nelsonIllustrisTNGSimulationsPublic2019,pillepichFirstResultsIllustrisTNG2018,
springelFirstResultsIllustrisTNG2018,nelsonFirstResultsIllustrisTNG2018,
naimanFirstResultsIllustrisTNG2018,marinacciFirstResultsIllustrisTNG2018},
a suite of cosmological hydrodynamic simulations carried out with the moving mesh 
code Arepo \citep{springelPurSiMuove2010}. 
Processes for galaxy formation, such as gas cooling, 
star formation, stellar feedback, metal enrichment, and AGN feedback, 
are simulated with subgrid prescriptions tuned 
to match a set of observational data
\citep[see][]{weinbergerSimulatingGalaxyFormation2017,
pillepichSimulatingGalaxyFormation2018}. 
A total of 100 snapshots, from redshift $z=20.0$ to $0$, are saved for 
each run. Halos are identified with the friends-of-friends (\specialname[FoF]) 
algorithm \citep{davisEvolutionLargescaleStructure1985} with a linking length of $0.2$, 
and subhalos are identified with the \specialname[Subfind] 
algorithm \citep{springelPopulatingClusterGalaxies2001,dolagSubstructuresHydrodynamicalCluster2009}.
Subhalo merger trees are constructed using the \specialname[SubLink]
algorithm \citep{rodriguez-gomezMergerRateGalaxies2015}.
To achieve a balance between sample size and resolution, we use the TNG100-1 
run (thereafter TNG). 

The second is TNG100-1-Dark, the DMO counterpart of TNG (thereafter TNG-Dark). 
It is run with the same cosmological parameters, box size, initial conditions, 
output snapshots as the hydrodynamic run, 
and it has mass and spatial resolutions similar to TNG.

The third is ELUCID \citep{wangELUCIDEXPLORINGLOCAL2016}, a DMO simulation 
obtained using the N-body code \specialname[L-GADGET], a memory optimized 
version of \specialname[GADGET-2] \citep{springelCosmologicalSimulationCode2005}. 
A total of 100 snapshots, from redshift $z=18.4$
to $0$, are saved. Halos, subhalos and subhalo merger trees are identified 
and constructed using the same algorithms as TNG. 

The final one is EAGLE 
\citep{schayeEAGLEProjectSimulating2015, 
crainEAGLESimulationsGalaxy2015,mcalpineEAGLESimulationsGalaxy2016,
theeagleteamEAGLESimulationsGalaxy2017}, a suite of cosmological 
hydrodynamic simulations run with the \specialname[GADGET-3] tree-SPH code,
an extension of \specialname[GADGET-2] \cite{springelCosmologicalSimulationCode2005}. 
A total of 29 snapshots, from $z=20$ to $0$, are saved. Halos and subhalos are 
identified by the same algorithms as TNG. Subhalo merger trees 
are constructed by the \specialname[D-Trees] algorithm \citep{jiangNbodyDarkMatter2014}.
We use the high-resolution run, EAGLE Ref-L0100N1504 (thereafter EAGLE) for
our analysis, which has a resolution comparable to TNG. 

The cosmology and simulation parameters of all the four simulations 
are listed in Table~\ref{tab:simulations}.

\subsection{Subhalo Properties}

The subhalo catalogs of both TNG and EAGLE present a variety of quantities, such as stellar mass, 
halo mass and star formation rate. The subhalo catalog of ELUCID gives the properties of
of dark matter halos. 
As demonstrated by \cite{chenRelatingStructureDark2020}, halo properties themselves 
have significant degeneracy, so that it is not necessary to include all halo properties 
in an empirical model. Furthermore, as demonstrated in Paper-I, 
galaxy stellar properties depend significantly only on a subset of all the halo 
properties. Motivated by the results of these two papers, we choose to use the following 
set of subhalo properties that are the most relevant to the halo-galaxy interconnection.
\bit
\item $\Mhalo$: the `top-hat' mass of the host FoF halo of a subhalo. 
	This halo mass is calculated within a virial radius within which the overdensity 
	is equal to that given by the spherical collapse model 
	\citep{bryanStatisticalPropertiesXRay1998}. 
	The corresponding virial radius and virial velocity are denoted as 
	$\Rhalo$ and $\Vhalo$, respectively.
\item $j_{\rm infall}$: the normalized orbital angular momentum of a 
	satellite subhalo, defined as \beq 
		j_{\rm infall} = 
			\frac{\|\Delta {\bf r} \times \Delta {\bf v}\| }{ \sqrt{2}R_{\rm h, cent}V_{\rm h, cent}} ,
	\eeq 
	where $\Delta {\bf r}$ and $\Delta {\bf v}$ are the position and velocity 
	of the satellite relative to its central subhalo, respectively, and
	$R_{\rm h, cent}$ and $V_{\rm h, cent}$ are the virial radius and virial velocity 
	of the central subhalo, respectively
	\footnote{
		To avoid confusion, we use $\| \|$ to represent the vector 2-norm or 
		matrix Frobenius norm, and $\log$ to denote 10-based logarithm. We use 
		1,2, and 3-$\sigma$ to denote 
		regions covering $68\%$, $95\%$ and $99.7\%$ of the data points, 
		respectively.
	}.
	$j_{\rm infall}$ is defined for a satellite subhalo at the infall time.
\item $\tau_{\rm merge}$: the merger time of a satellite subhalo, defined as 
	\beq 
		\tau_{\rm merge} = \log \frac{1+\zmerge}{1+\zinfall},
	\eeq
	where $\zmerge$ is the redshift just before the satellite merges 
	into the central subhalo, and $\zinfall$ is the redshift when  
	the satellite falls into the host FoF halo. 
\item $I_{\rm merge}$: a binary indicator to describe whether or not a subhalo has merged
	by $z=0$.
	If it has merged, $I_{\rm merge}=1$; otherwise $I_{\rm merge}=0$.
	Note that $\tau_{\rm merge}$ is undefined 
	for satellites that have not yet merged by $z=0$. 
	$I_{\rm merge}$ is defined to include such satellites in
	our model (\S\ref{ssec:model-of-satellite}).
\item $\Mstar$: the stellar mass of a subhalo. This is defined as the mass within twice 
	the stellar half mass radius for 
	TNG, and within $30$ physical kpc for EAGLE.
\item $\rm SFR$: the star formation rate within the same radius as that for $\Mstar$.
\item $\MstarInt$: the total stellar mass ever formed in the history of a subhalo: 
	$\sum_n {\rm SFR}_n \Delta t_n$, where ${\rm SFR}_n$ is the SFR at $n$th snapshot 
	in the history and $\Delta t_n$ is the time interval spanned by the snapshot. 
	$\MstarInt$ will be the direct output of our model in \S\ref{sec:model}.
	It is different from $M_*$ in that the mass loss due to 
	stellar evolution and mass changes due to mergers are not taken into account.
	Merger-triggered changes in the SFR are included in $\MstarInt$.
	When comparing with observations, these effects should be included 
	by properly assuming a stellar evolution model, an initial 
	mass function \citep[e.g.,][]{salpeterLuminosityFunctionStellar1955,
	chabrierGalacticStellarSubstellar2003,zhouSDSSIVMaNGAStellar2019}, 
	and a merger model. Our test using the TNG simulation shows that the simple
	addition of $\MstarInt$, of a galaxy with all the progenitors merged 
	into it, is $\sim 0.3\ {\rm dex}$ larger than $\Mstar$ at all redshifts 
	for $\Mstar > 10^8 \msun$.

\item sSFR: the specific star formation rate, defined as 
	${\rm sSFR} = {\rm SFR} / \MstarInt$.
\eit 
Note that $\Mstar$, SFR, $\MstarInt$ and sSFR are defined only in the 
hydrodynamic simulations, TNG and EAGLE. Other properties are defined also in 
TNG-Dark and ELUCID.

As shown in Paper-I, the halo-galaxy connection depends not only on the current 
status of a halo, but also on its assembly history. 
Following Paper-I, we define the subhalo mass assembly history (MAH), $\MhaloVec$,
for a central subhalo as the set of $M_{\rm h}$ values in the main branch of 
the subhalo merger tree rooted in this subhalo. 
We also define the star formation history (SFH), $\MstarIntVec$, 
of a galaxy as the set of $\MstarInt$ values in the main 
branch of the subhalo merger tree rooted in the host subhalo of the galaxy. 

The discrete forms of the MAH and the SFH in general are each a `vector' (tuple) 
in high-dimensional configuration space, and 
the information contained in these vectors may be highly degenerate. 
To extract useful information from them, some method of 
dimension reduction is needed. 
Here we follow Paper-I 
(see its Appendix A for a detailed description) 
to reduce the dimensionality of the MAH and the SFH by \specialname[PCA] whenever it is needed. 
The application of \specialname[PCA] to a MAH (or a SFH) reduces it to  
a set of PCs, with the first several expected to be capable of 
capturing its main properties. We denote the PCs of the MAH and 
SFH as ${\bf pc}_{\rm h}$ and ${\bf pc}_{*}$, respectively. 
The details of the related analyses are described in \S\ref{sec:model}.

\subsection{Tree Decomposition and Subhalo Samples}
\label{ssec:samples}

To reduce the complexity of the empirical model, we do not attempt to model 
the stellar content for each single subhalo. Instead, we decompose each subhalo 
merger tree into a set of disjoint branches, $\{b\}$, each of which is a 
chain of subhalos that form the main branch of a root subhalo.
The decomposition is processed through the following steps: 
\bit 
	\item Starting from the root subhalo $h_{r}$  of the whole subhalo merger tree 
		(i.e., the subhalo that does not 
		have any descendant), we use all subhalos in the main branch of $h_r$ 
		to form a single branch $b$. 
	\item Subhalos attached to  $b$ are removed from the tree, resulting in a set of 
		sub-trees of the original tree.
	\item Treating each sub-tree as a new `tree', 
	    we recursively perform the same decomposition for all the  
	    sub-trees until all subhalos in the original tree 
		are assigned into branches. All the branches collectively form 
		$\{b\}$.
\eit 
For each branch, we walk through it from high to low redshift. 
We define the infall redshift $\zinfall$ of the branch as the 
redshift of the last snapshot when the subhalo is still a central subhalo. 
We define the infall halo mass $M_{\rm h, infall}$ as the halo mass at $\zinfall$. 
Note that this definition is valid for branches in both sub-trees and the original tree. 

To achieve a balance of numerical stability and sample size, we select 
all branches with $M_{\rm h, infall}>M_{\rm h, limit} = 10^{11} \msun$. 
The subhalos in the selected branches are the main sample we use
in our model and analysis.
Such a selection ensures that galaxies are well resolved in TNG and EAGLE, 
and that the sample size is still sufficiently large to allow for model 
learning and testing. 
We have checked that our model is stable when using a lower mass limit.
To alleviate computational cost, we apply and test our model 
using a sub-box with $200 \mpc$ side-length from ELUCID. 
As shown in Appendix~\ref{app:subhalo-history} and Figure~\ref{fig:app-hmf},
this sub-box can already significantly reduce the cosmic variance 
in comparison with TNG and EAGLE, and is thus sufficient for our purpose.

The above mass limit is suitable for TNG, TNG-Dark and EAGLE, 
but is too small for ELUCID. The lowest halo mass that can be resolved by ELUCID is 
$\sim 6 \times 10^9 \msun$, about $50 \times$ larger than that in the other three simulations. 
Because of this, the ELUCID simulation cannot trace the MAH of a 
subhalo to sufficiently high redshift when the star formation in 
the branch is already significant.
To overcome this limitation, we use merger trees from TNG-Dark
to extend all branches in ELUCID down to the same mass limit. 
For each branch in ELUCID, we pick a branch in TNG-Dark at 
the same $\zinfall$ and with the same $M_{\rm h,infall}$. 
The missed part of MAH in ELUCID at high redshift is extended by this 
picked branch, with proper interpolation to adjust the 
redshift sampling.
Note that this is different from \cite{chenELUCIDVICosmic2019} 
where analytical halo merger trees obtained from a Monte Carlo 
(MC) implementation were used to extend the MAH. Our choice is 
motivated by the fact that the merger trees from high-resolution simulations are  
more precise, and are usually used to calibrate the analytical trees.
As shown in Appendix~\ref{app:subhalo-history}, the extended MAHs of ELUCID 
match well with those obtained from the other three simulations.
Any branch that terminates before it reaches the upper redshift limit
(set by the redshift of the first snapshot) 
is padded with a small value for numerical stability.

\section{The Empirical Model}
\label{sec:model}

\subsection{Overall Design Strategies}
\label{ssec:overall-design}
\begin{figure*} \centering
	\includegraphics[width=\textwidth]{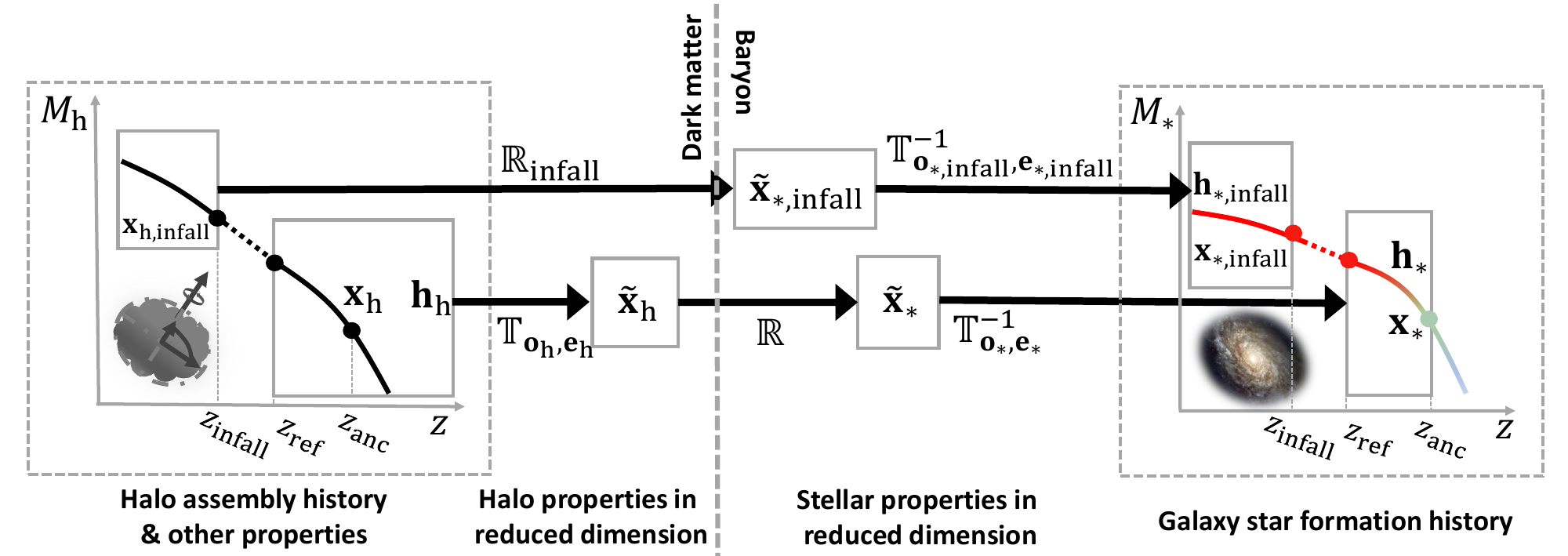}
	\caption{The pipeline of the model \specialname[MAHGIC]. 
		The model separately deals with 
		central (\textbf{lower channel} in the figure) 
		and satellite galaxies (\textbf{upper channel}).
		Galaxy properties at the anchor redshift, ${\bf x}_{*}$, 
		and galaxy SFH in the central stage, ${\bf h}_*$, are mapped from subhalo MAH 
		${\bf h}_{\rm h}$, and other halo properties 
		at the anchor redshift, ${\bf x}_{\rm h}$. Three transformations
		$(\mathbb{T}_{{\bf o}_{\rm h}, {\bf e}_{\rm h}}, 
		\mathbb{R}, \mathbb{T}_{ {\bf o}_*, {\bf e}_* })$ 
		are trained and applied for the central stage.
		Galaxy properties at the infall time, ${\bf x}_{\rm *,infall}$,
		and galaxy SFH in the satellite stage, ${\bf h}_{\rm *, infall}$, 
		are mapped from halo properties 
		at the infall time, ${\bf x}_{\rm h,infall}$. Two transformations 
		$(\mathbb{R}_{\rm infall}, \mathbb{T}_{{\bf o}_{\rm *, infall}, {\bf e}_{\rm *, infall}})$ 
		are trained and applied to the satellite stage. The whole pipeline
		consists of multiple redshift pieces, one of which is shown here. }
	\label{fig:model-outline}
\end{figure*}
\begin{center}
\begin{table}
\caption{Notations for variables and transformations used by \specialname[MAHGIC].
Notations for central galaxies are listed here, 
while those for satellite galaxies can be obtained by adding a subscript 
`infall' (e.g., ${\bf x}_{\rm h, infall}$ for 
the properties of satellite subhalo at infall time). The order of the listed notations 
follows the pipeline in the application phase (see \S\ref{sec:model}), which 
is consistent with information flow in Figure~\ref{fig:model-outline}.}
\begin{tabularx}{\columnwidth}{ c | >{\hsize=.82\hsize}X }
	\hline 
	Variables 			& 				Explanation 				 \\ 
	\hline \hline
	$({\bf x}_{\rm h}, {\bf h}_{\rm h})$	      
		& Physical properties of subhalo. ${\bf x}_{\rm h}$ are subhalo properties at 
		  the anchor redshift $z_{\rm anc}$
		  of the tree branch. ${\bf h}_{\rm h}$ is the MAH of this branch. \\ \hline
	
	${\bf \tilde{x}}_{\rm h}$
		& Subhalo properties in the space of reduced dimension, including 
		  ${\bf x}_{\rm h}$ and a set of PCs of subhalo MAH. It is connected 
		  to the physical subhalo properties through the representation 
		  transformation of subhalo, 
		  ${\bf \tilde{x}}_{\rm h} = \mathbb{T}_{{\bf o}_{\rm h}, {\bf e}_{\rm h}} ({\bf x}_{\rm h}, {\bf h}_{\rm h}) $.
		  \\ \hline 	

	${\bf \tilde{x}}_*$
		& Galaxy properties in the space of reduced dimension, including 
		${\bf x}_*$ and a set of PCs of galaxy SFH. It is produced through the 
		halo-galaxy mapping, ${\bf \tilde{x}}_* = \mathbb{R}({\bf \tilde{x}}_{\rm h})$.
		\\ \hline 
	
	$({\bf x}_*, {\bf h}_*)$
		& Physical properties of galaxy. ${\bf x}_*$ are galaxy properties at $z_{\rm anc}$. 
		${\bf h}_*$ is the SFH of galaxies 
		in this branch.
		They are produced through the inverse of the representation transformation of galaxy, 
		$({\bf x}_*, {\bf h}_*) = \mathbb{T}^{-1}_{ {\bf o}_*, {\bf e}_* }({\bf \tilde{x}}_*)$.
		\\ \hline
\end{tabularx}
\label{tab:variables}
\end{table}
\end{center}

As discussed in \S\ref{sec:intro}, the goal of \specialname[MAHGIC] 
is to first use hydrodynamic simulations to motivate and train our model design, 
and then to apply it to DMO simulations. Motivated
by the results obtained in \cite{chenRelatingStructureDark2020} and Paper-I, 
we adopt the following strategies to construct the model:
\benum
\item Central and satellite galaxies are modeled separately. This is motivated 
	by the fact that processes regulating star formation  
	are very 
	different for the two populations. For example, a satellite galaxy after infall 
	may undergo significant environmental quenching due to tidal stripping and 
	ram-pressure stripping, which may be less important for a central galaxy. 
	Such a separation is commonly adopted in other empirical models 
	\citep[e.g.,][]{yangEVOLUTIONGALAXYDARK2012,
	mutchSimplestModelGalaxy2013,
	luEmpiricalModelStar2014,luStarFormationStellar2015, 
	hearinIntroducingDecoratedHODs2016, mosterEmergeEmpiricalModel2018,
	behrooziUniverseMachineCorrelationGalaxy2019a}.

\item Halo properties used in the model are required to be robust. They should  
	be insensitive to baryonic effects and stable against changes in numerical resolution. 
	This is required by our goal, as we want to train our model using hydrodynamic 
	simulations, which contain baryonic effects and usually have a high resolution,  
	and apply it to large DMO simulations where the baryonic effects
	are absent and the numerical resolution may be different. To reduce the 
	impact of baryonic and resolution effects, we use the assembly history 
	represented by halo mass $\Mhalo (z)$ as the main 
	predictor of stellar properties, and avoid using the assembly history 
	represented by the maximum circular velocity, $v_{\rm max}(z)$, 
	which is sensitive to both effects.   
	For the model of satellite galaxies, we use  
	halo properties at the infall time as predictors, and avoid halo properties 
	after the infall. As tested by us using TNG and ELUCID simulations,
	halo properties after infall are sensitive to baryonic effects and 
	numerical resolution, while quantities defined at the infall time are 
	stable (see Appendix~\ref{app:subhalo-history}). 
	Due to limited resolutions, large-volume DMO simulations like ELUCID in general  
	are incapable of fully tracing the mass assembly history (MAH) of a halo to high 
	redshift when its main progenitor becomes too small to be resolved. 
	For such cases, we use the method 
	described in \S\ref{ssec:samples} to extend the MAH down to a sufficiently low 
	mass limit.

\item The model must be able to capture potential non-linearities in the 
	halo-galaxy interconnection. We, therefore, build a deep model with multiple 
	layers including representation, halo-galaxy mapping, and 
	reconstruction.
	This mimics modern deep neural networks, where
	input values are first transformed to a simple representation and 
	then fed into a traditional regressor or classifier to produce the output.

\item The model must be interpretable. This is needed because we want to understand 
	the physics underlying the model, rather than just building a 
	`black-box' model. We achieve this by using interpretable prescriptions 
	in all the layers of the model and we optimize them separately, an approach 
	similar to
	the greedy algorithm in algorithm-design 
	\citep[e.g.][]{cormenIntroductionAlgorithmsThird2009,sedgewickAlgorithms2011}. 
	To be specific, in the representation and reconstruction layers, 
	we use \specialname[PCA] to 
	reduce the dimensionality of the subhalo MAH and the galaxy SFH. 
	\specialname[PCA] is a simple and yet powerful 
	dimension-reduction technique with robust mathematical 
	interpretability. As demonstrated by \cite{chenRelatingStructureDark2020}, 
	the principal components (PCs) of subhalo MAH carry sufficient information about how 
	halos form and are also strongly correlated with other halo structural 
	and environmental properties. In addition, as demonstrated in Paper-I, 
	these PCs are strongly correlated with galaxy SFH, thus providing
	an ideal way to do the halo-galaxy mapping. In the layer of 
	halo-galaxy mapping, we adopt decision tree classifiers and regressors
	to map halo properties to galaxy stellar properties. Tree-based 
	models are non-linear, so they are capable of dealing with
	potential non-linearities in the model.
	Trees are also interpretable through the importance values 
	$\mathcal{I}(x)$ of individual predictors and the $R^2$ value 
	of model performance (see Paper-I). Finally, 
	the predicted stellar properties are used in the reconstruction 
	layer to obtain the physical SFH.

\item The model should be flexible enough to accommodate constraints 
	from current and future observations, and yet avoid over-fitting.
	In the context of Bayesian inference, model complexities 
	can be increased to capture more subtle processes
	as more observational constraints become available. 
	For the problem of galaxy formation concerned here, 
	constraints are obtained from observations of galaxies at different 
	redshifts. Thus, it is not useful to build a mapping that is 
	valid only at a given redshift.
	Instead, we should construct the mapping on the basis of 
	MAH and SFH. To do this, we use tree branches of MAH and SFH 
	described in \S\ref{ssec:samples}	
	as individual entries, and build a mapping of the PCs 
	between the MAH and SFH branches. 
	This has the advantage of avoiding an over-complicated model.
	We use \specialname[GBDT] in the halo-galaxy mapping to suppress over-fitting, 
	as described in detail in Paper-I (see its Appendix B). 
	As the amount of constraining data increases, one can use more PCs 
	and more halo quantities as input features to accommodate 
	the additional constraints. 
\eenum

{\revisestyle  
To improve the accuracy of our model, we need to break the modeling 
into multiple pieces in redshift. The reasons for this are the following.
First, the prediction of stellar properties is the most precise 
at the redshift where predictor halo properties are used. 
Indeed, as shown in Paper-I, $\MstarInt$ at a given redshift
is predominantly determined by $\Mhalo$ at the redshift.
The extension of the SFH towards higher redshift in general introduces 
cumulative errors, even if PCs are used, because the predictions of stellar PCs 
themselves contain error, as seen from Figure~\ref{fig:var-contrib-cent}.
Second, branches of different $\zinfall$ have different lengths in redshift 
coverage. To make the PCA feasible, we need to align the branches so that 
that they have the same length
(see \S\ref{ssec:model-of-central} and \S\ref{ssec:model-of-satellite}). 
If only one piece is used, artificial continuations are needed for 
MAHs that have different lengths, which can introduce artifacts and 
should be avoided.

Using multiple pieces is an analog to a non-parametric approach: 
it solves the problem at the expense of the interpretability. 
However, our model can avoid problems in a full non-parametric 
approach by using PCs for each of the pieces. In principle,  
the number of pieces and the locations of the breaking redshift 
can both be set as free hyperparameters and adjusted by 
an independent cross-validation process, depending on the quality of 
the training data and the form of the objective function.
This is similar to a neural network, where neurons in the same layer are 
responsible for different regions of the feature space and should be 
adjusted according to the training data.
}

Taking account of all these requirements, we intend to
build a deep interpretable model for galaxy 
formation in dark matter halos. 
The details are described in the following sections separately 
for central galaxies (\S\ref{ssec:model-of-central}) and satellite galaxies 
(\S\ref{ssec:model-of-satellite}). Figure~\ref{fig:model-outline} shows 
the outline of the model, and Table~\ref{tab:variables} lists the 
variables and transformations involved.

\subsection{The Model for Central Galaxies}
\label{ssec:model-of-central}

Our model for central galaxies follows closely that of Paper-I, 
with some modifications. In Paper-I we only modeled galaxies that 
are centrals at $z=0$ ($\zinfall =0$ according to our notation in this paper).
Here we extend the modeling to include subhalo branches with all 
$\zinfall$. The procedures of our model are slightly different 
between the training phase and the application phase. 
For the training, we have information about both subhalos 
and galaxies, while for the application only subhalo properties 
are accessible. We describe the modeling in the training phase
first, and then highlight the changes in the application phase.

{\bf Training phase.}
The goal of our model for central galaxies is to populate each 
branch with galaxies for all subhalos in the branch 
at $z\ge \zinfall$. Because the halo-galaxy relation is expected 
to change with redshift, we break the whole redshift range into $N_{\rm piece}$ 
pieces with separation redshift at 
$(z_0, z_1, z_2, ..., z_{N_{\rm piece}})$, where $z_0 = 0$ and 
$z_{N_{\rm piece}}$ is chosen to be sufficiently high to cover the desired 
redshift range. The model is built independently for each 
piece, with the $i$th piece responsible for all central galaxies 
in the redshift interval $(z_{i-1}, z_i]$. For the sake of 
description, we refer to $z_{i-1}$ as the reference redshift and 
denote it as $\zrefer$; and to $z_{\rm i}$ as the anchor redshift 
and denote it as $\zanc$. The relevance of these two redshifts 
to our description will become clear later in this section.

To model the $i$th piece, we select a reference sample, 
${\rm S}_{\rm ref}$, which is defined as
all tree branches with $\zinfall=\zrefer$.
For each branch in ${\rm S}_{\rm ref}$, 
the subhalo MAH at $z < \zinfall$ is cut out, 
and the remaining MAH is denoted as $\bf{h}_{\rm h}$, 
which is a vector representing the values of $\Mhalo$ for all subhalos in this branch. 
We also take a set of halo properties at $z=\zanc$, 
and denote them as ${\bf x}_{\rm h}$. We normalize ${\bf h}_h$ as 
\beq 
{\bf \tilde{h}}_{\rm h} = \log \frac 
		{{\bf h}_{\rm h}}
		{M_{{\rm h},z=\zanc}},
\label{eq:MAH-normalization}
\eeq 
and we apply \specialname[PCA] to ${\bf \tilde{h}}_{\rm h}$ to 
obtain a set of PCs, ${\bf pc}_{\rm h}$, a 
mean MAH ${\bf o}_{\rm h}$, and a set of eigen 
modes ${\bf e}_{\rm h}$. These PCs are combined with the 
set of halo properties, ${\bf x}_{\rm h}$, to form the vector 
${\bf \tilde{x}}_{\rm h} = ({\bf x}_{\rm h}, {\bf pc}_{\rm h})$. 
This vector is the output of the representation layer, and 
is to be fed into the halo-galaxy mapping layer. 

In this paper, we use $\Mhalo$ as the only properties at $\zanc$, 
i.e., ${\bf x}_{\rm h}=(\log M_{{\rm h},z=\zanc})$. 
As shown in Paper-I, the halo mass at a given redshift 
is the dominating factor in determining the stellar properties 
of the central galaxy hosted by the halo at the same redshift. 
For the output of the \specialname[PCA], we follow Paper-I and 
use the first two PCs. More quantities and PCs can be added into 
${\bf \tilde{x}}_{\rm h}$ when needed. 

After the construction of ${\rm S}_{\rm ref}$ and the \specialname[PCA] 
template $({\bf o}_{\rm h}, {\bf e}_{\rm h})$, 
we select all the remaining branches with $\zinfall \leqslant \zanc$ that are not 
included in the reference sample. For any of these branches, 
if $\zinfall < \zrefer$, subhalos with $z < \zrefer$ are cut out from the 
branch. Otherwise, if $\zinfall > \zrefer$, satellite subhalos 
with $z < \zinfall$ are cut out, 
and the missed history between $\zrefer$ and $\zinfall$ is 
completed using the mean assembly rate of the reference sample
scaled by the halo mass at $\zinfall$. 
With the trimming and completion, all MAHs in the remaining sample 
are vectors of the same length. They are then normalized and transformed 
by the \specialname[PCA] template $({\bf o}_{\rm h}, {\bf e}_{\rm h})$ obtained  
from ${\rm S}_{\rm ref}$ to yield ${\bf \tilde{x}}_{\rm h}$ for these 
branches. The whole process of representation transformation 
from $({\bf x}_{\rm h}, {\bf h}_{\rm h})$ to 
${\bf \tilde{x}}_{\rm h}$ is denoted as 
$\mathbb{T}_{{\bf o}_{\rm h}, {\bf e}_{\rm h}}$, and we write
\beq
	{\bf \tilde{x}}_{\rm h} = \mathbb{T}_{{\bf o}_{\rm h}, 
		{\bf e}_{\rm h}}({\bf x}_{\rm h}, {\bf h}_{\rm h}).
\eeq 

We use the same technique to build the representation of the galaxy SFH in reduced 
dimensions. The differences are that we replace all halo quantities with 
galaxy stellar properties, and that we use only SFH at $z \leqslant \zanc$ because 
they are the ones relevant for the piece in question. 
Here we define ${\bf x}_* = (\log M_{*, \rm int, z=\zanc})$ 
to be the set of stellar properties at $\zanc$; ${\bf h}_*$ to 
be the SFH described by $M_{\rm *, int} (z)$ in the branch, with the 
same trimming and completion steps as those applied to the halo MAH; 
${\bf \tilde{h}}_*$ to be the SFH normalized by $M_{*, \rm int, z=\zanc}$ 
using the same technique as in Eq.~\ref{eq:MAH-normalization}; 
and ${\bf pc}_*$ to be the first two PCs of ${\bf \tilde{h}}_*$. 
The transformation $\mathbb{T}_{ {\bf o}_*, {\bf e}_* }$
is also built using ${\rm S}_{\rm ref}$. Denoting the stellar properties in the 
space of reduced dimension by 
${\bf \tilde{x}}_* = ({\bf x}_*,{\bf pc}_*)$, we can summarize the 
whole process by 
\beq 
{\bf \tilde{x}}_* = \mathbb{T}_{ {\bf o}_*, {\bf e}_* } ( {\bf x}_*, {\bf h}_* ).
\eeq 

Represented by ${\bf \tilde{x}}_{\rm h}$ and ${\bf \tilde{x}}_*$,
the dimensions of MAH and SFH are significantly reduced, 
which makes it possible to construct a interpretable 
non-linear mapping between them. Here we adopt a \specialname[GBDT] $\mathbb{R}$ 
to map ${\bf \tilde{x}}_{\rm h}$ to ${\bf \tilde{x}}_*$:
\beq
{\bf \tilde{x}}_* = \mathbb{R}({\bf \tilde{x}}_{\rm h}).
\eeq 
This mapping is trained by ${\bf \tilde{x}}_{\rm h}$ and ${\bf \tilde{x}}_*$ 
obtained from the hydrodynamic simulation that is used for the training.

Because a part of the galaxies show rapid quenching at low $z$ 
(see, e.g., Figure~\ref{fig:bimodality}), we follow Paper-I 
to separate the first piece of our central model into two, one 
for star-forming galaxies (${\rm sSFR}_{z=0} \geqslant 10^{-2} \gyri$) 
and the other for quenched galaxies (${\rm sSFR}_{z=0} < 10^{-2} \gyri$). 
The set of subhalo properties, ${\bf \tilde{x}}_{\rm h}$, is used 
to classify the branch as star-forming or quenched, 
and the branch is then sent into the corresponding pipeline.

{\bf Application phase. } 
In the application of the model to a test simulation, 
the pipeline goes in a different direction,
because now only halo properties, 
$({\bf x}_{\rm h}, {\bf h}_{\rm h})$, are available to us.
The prediction of the galaxy SFH from halo properties 
is achieved through three consecutive transformations, 
\beq 
	( {\bf x}_*, {\bf h}_* ) =
	\mathbb{T}^{-1}_{ {\bf o}_*, {\bf e}_* } 
	\mathbb{R}
	\mathbb{T}_{{\bf o}_{\rm h}, {\bf e}_{\rm h}}
 	({\bf x}_{\rm h}, {\bf h}_{\rm h}),
\eeq
where $\mathbb{T}_{{\bf o}_{\rm h}, {\bf e}_{\rm h}}$ is obtained from 
${\rm S}_{\rm ref}$ of the test simulation, while $\mathbb{R}$ 
and $\mathbb{T}_{ {\bf o}_*, {\bf e}_* }$ are obtained
in the training phase. Finally,  
the piece of $\bf{h}_*$ in the redshift range 
$(\max(\zrefer, \zinfall), \zanc]$ is the model output of the $i$th piece 
for the branch in question. After all the pieces are modeled for a branch, 
a proper smoothing is made at each of the separation redshifts 
to join the pieces together to form the whole stellar history of the branch.

\subsection{The Model for Satellite Galaxies}
\label{ssec:model-of-satellite}

For the reasons given in \S\ref{ssec:overall-design}, 
we only use subhalo properties at the infall time as input features 
for our model of satellite galaxies.  
We choose ${\bf x}_{\rm h, infall}=(M_{\rm h, infall}, M_{\rm h, infall, cent}, 
j_{\rm infall}, z_{\rm infall})$. 
Here $M_{\rm h, infall}$ is the subhalo mass of the target satellite,
$M_{\rm h, infall, cent}$ is the mass of its central subhalo, 
and $j_{\rm infall}$ is the normalized orbital angular momentum, 
all calculated at the infall time $z_{\rm infall}$.
This choice is motivated by results of dynamical friction studies 
\citep[e.g.,][]{boylan-kolchinDynamicalFrictionGalaxy2008a}, 
where the first three quantities are found to be the 
main factors affecting the orbital dynamics of a subhalo after infall.
The satellite model is also broken into $N_{\rm piece}$ pieces,
and each piece is responsible for a set of branches.  
The training and application phases of the model are described in the following.

{\bf Training phase.} 
The $i$th piece of our satellite model begins with selecting all branches with 
$\zinfall \in (z_{i-1}, z_{i}]$. The goal of the model is to populate these branches 
with galaxies at $z<\zinfall$. Because the halo properties ${\bf x}_{\rm h, infall}$
are already in low-dimension space, we only need to deal with dimension reduction 
for the galaxies.

Galaxy SFHs of the selected branches before $\zinfall$
are cut out because they have already been modeled by the central model. 
Since the SFHs of the branches with different $\zinfall$ may have different 
lengths , we pad them at the low-$z$ end with a constant $\MstarInt$ 
given by the last traceable snapshot, so as to make all SFHs 
have the same length. These SFHs are denoted as ${\bf h}_{\rm *, infall}$.
We also select a set of galactic
properties at the infall time, and denote them collectively as ${\bf x}_{\rm *,infall}$.
Thus, the input of the satellite model in the training phase is
$({\bf x}_{\rm *,infall}, {\bf h}_{\rm *, infall})$. To proceed further, 
we first normalize ${\bf h}_{\rm *, infall}$ using 
\beq
{\bf \tilde{h}}_{\rm *,infall} = \log \frac{{\bf h}_{\rm *, infall}}{M_{\rm *,int,z=\zinfall}},
\eeq
and then feed it into a \specialname[PCA] to obtain the PCs of the SFH, ${\bf pc}_{\rm *, infall}$,  
the mean SFH, ${\bf o}_{\rm *, infall}$, and a set of eigen modes, ${\bf e}_{\rm *, infall}$.
All the transformations on satellite galaxies are represented collectively 
by a single operator, $\mathbb{T}_{ {\bf o}_{\rm *, infall}, {\bf e}_{\rm *, infall}} $, 
which reduces satellite stellar properties to a vector in a space of reduced dimension, 
${\bf \tilde{x}}_{\rm *,infall}=({\bf x}_{\rm *,infall}, {\bf pc}_{\rm *, infall})$. 
Symbolically, we write
\beq 
	{\bf \tilde{x}}_{\rm *,infall} = 
		\mathbb{T}_{{\bf o}_{\rm *, infall}, {\bf e}_{\rm *, infall}}
		({\bf x}_{\rm *,infall}, {\bf h}_{\rm *,infall}).
\eeq 

We choose ${\bf x}_{\rm *,infall}=(I_{\rm merge}, \tau_{\rm merge})$. 
This is different from the central model where $M_{\rm *,int,z=\zanc}$ is used. 
The reason for not including stellar mass is that 
the satellite model does not need an extra normalization 
in the reconstruction of SFH, because it is already provided by 
$M_{\rm *,int, infall}$ as an output of the central model. 
The reason for including merging variables is the following. 
In the application to DMO simulations of relatively low 
resolution, a subhalo after infall may not be robustly resolved
and may be destroyed artificially before merging into the central subhalo,
as demonstrated in Appendix~\ref{app:subhalo-history}.
Thus, when applying the model to a DMO simulation of low resolution,
such as ELUCID, we need to first predict the merging time correctly, 
and then extend the subhalo branch to the correct merging time. 
Note that this will produce some galaxies that do not have 
simulated subhalos associated with them. 

Now that both subhalos and galaxies are represented with reduced dimensions, 
we can move ahead to process the halo-galaxy mapping. As for the central model, 
we use \specialname[GBDT] learners to map halo properties to galaxy properties. 
The three learners used are: 
\bit 
	\item a regressor that maps halo infall properties
		${\bf x}_{\rm h, infall}$ to the SFH in PC space, ${\bf pc}_{\rm *,infall}$;
	\item a classifier that maps ${\bf x}_{\rm h, infall}$ to $I_{\rm merge}$; 
	\item a regressor that maps ${\bf x}_{\rm h, infall}$ to the merging time $\tau_{\rm merge}$ 
		of a branch if it is terminated/destroyed.
\eit 
These three learners are collectively treated as a single operator, so that 
\beq 
{\bf \tilde{x}}_{\rm *,infall} = \mathbb{R}_{\rm infall}({\bf x}_{\rm h, infall}).
\eeq 
All the learners are trained using the data from the training simulation.
As described above, the last two learners are useful when we apply our model 
to DMO simulations with resolutions lower than the training simulation. 

{\bf Application phase. } In the application of our model to a test simulation, 
only ${\bf x}_{\rm h, infall}$ is available, and we predict the stellar properties 
using 
\beq 
	({\bf x}_{\rm *,infall}, {\bf h}_{\rm *,infall}) =
	\mathbb{T}^{-1}_{{\bf o}_{\rm *, infall}, {\bf e}_{\rm *, infall}} 
	\mathbb{R}_{\rm infall} 
	({\bf x}_{\rm h, infall}),
\eeq 
with all the operators trained in the training phase.

Once $\MstarInt$ is modeled for both central and satellite galaxies, 
the corresponding SFR can be obtained by differentiating it along 
each branch. When applying the model to a DMO simulation with snapshots 
at redshifts different from that of the training simulation, 
interpolations are applied to the output SFH for the DMO simulation 
to adjust the redshift sampling.

As demonstrated in Paper-I (see their \S 3.1), the main-sequence 
scatter of the sSFR-$\Mstar$ relation cannot be fully explained 
by halo properties. By using a large set of halo properties, the 
explained scatter, as described by $R^2$ of their regressor, is 
still less than $50\%$ at both $z=0$ and $z=2$. Hence, a model 
that relies on halo properties, without over-fitting, 
always underestimates the scatter of sSFR for the main-sequence 
galaxy population. The missed scatter of $\log {\rm sSFR}$ 
in our model can be added as a normal random component 
whose variance is taken from the difference between the model output 
and the hydrodynamic simulation used as the training data.
{\revisestyle If required, this missing variance can also be 
modeled as a function of halo properties using an independent regressor.
This is critical when the model is to be extended to using 
observations as input, where only summary statistics are available.
}

The direction of information flow in our model is shown clearly 
by Figure~\ref{fig:model-outline}.
In the training phase of both the central and satellite models, 
information flows from the two ends to the center of the model pipeline,
while in the application phase, information flows in a single direction, 
from the left to the right. This is a direct outcome of our 
optimization strategy, and is different from a neural network-based 
deep model, where the information flows cyclically in the training phase 
if a gradient back-propagation algorithm \citep[e.g.,][]{rumelhartLearningRepresentationsBackpropagating1986} 
is adopted.

\begin{figure} \centering
	\includegraphics[width=\columnwidth]{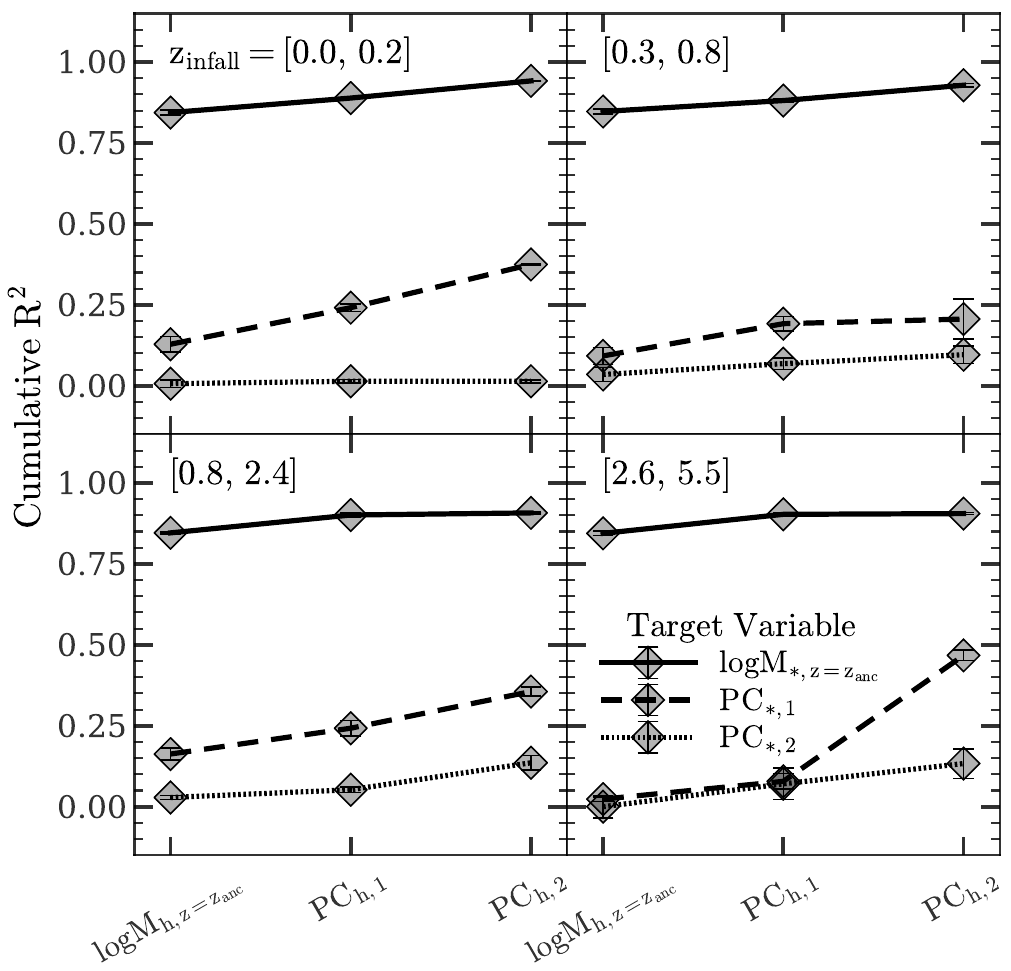}
	\caption{Cumulative $R^2$ of the regressor $\mathbb{R}$ 
		in the model of central galaxies trained by TNG. Each 
		panel shows the result for one piece of the model whose 
		infall redshift range is indicated at the upper left corner 
		of the panel. Error bars are estimated by 
		the resampling method described in \S\ref{ssec:var-contrib}. }
	\label{fig:var-contrib-cent}
\end{figure}

\begin{figure} \centering
	\includegraphics[width=7cm]{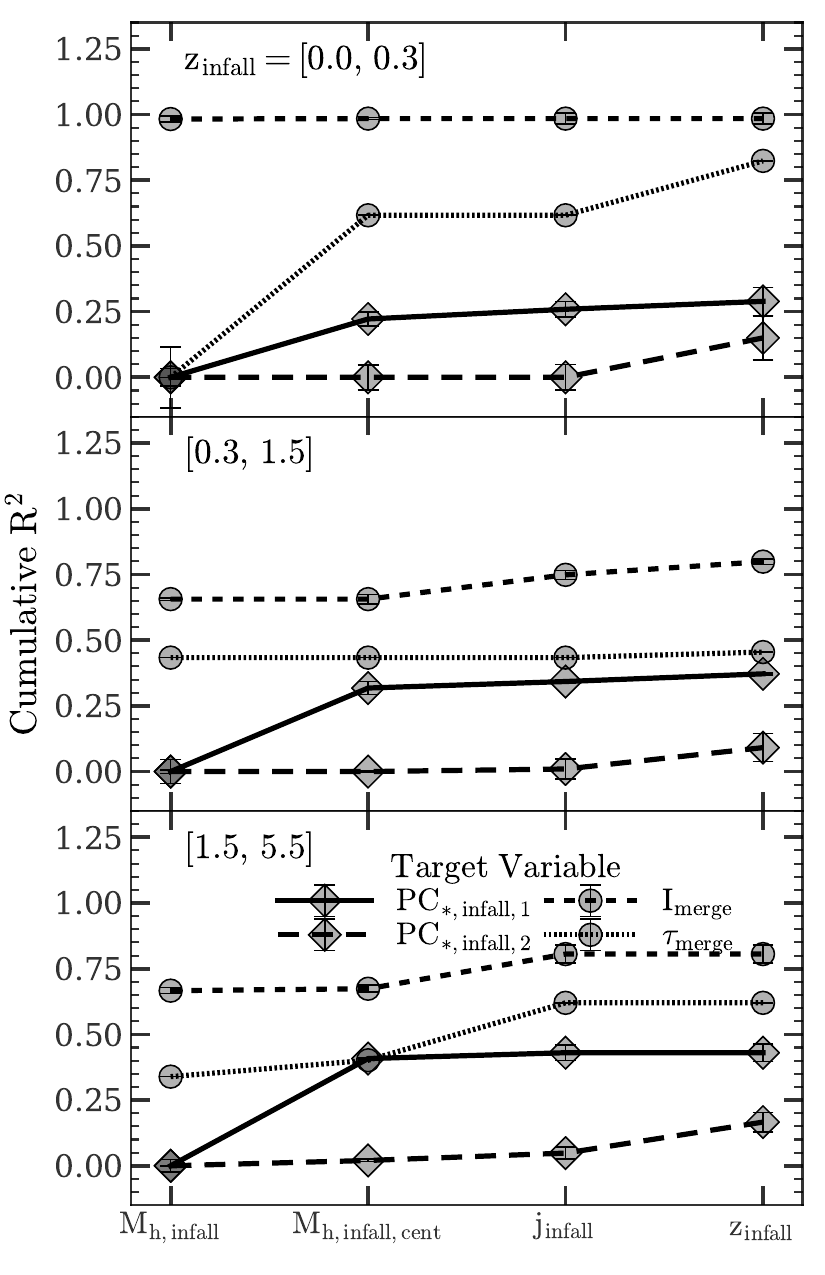}
	\caption{Cumulative $R^2$ of the mapping $\mathbb{R}_{\rm infall}$ 
		in the model of satellite galaxies trained by TNG. Each 
		panel shows the result for one piece of the model whose 
		infall redshift range is indicated at the upper left corner 
		of the panel. 
		Error bars are estimated by 
		the resampling method described in \S\ref{ssec:var-contrib}. 
		}
	\label{fig:var-contrib-sat}
\end{figure}

\begin{figure*} \centering
	\includegraphics[width=13.5cm]{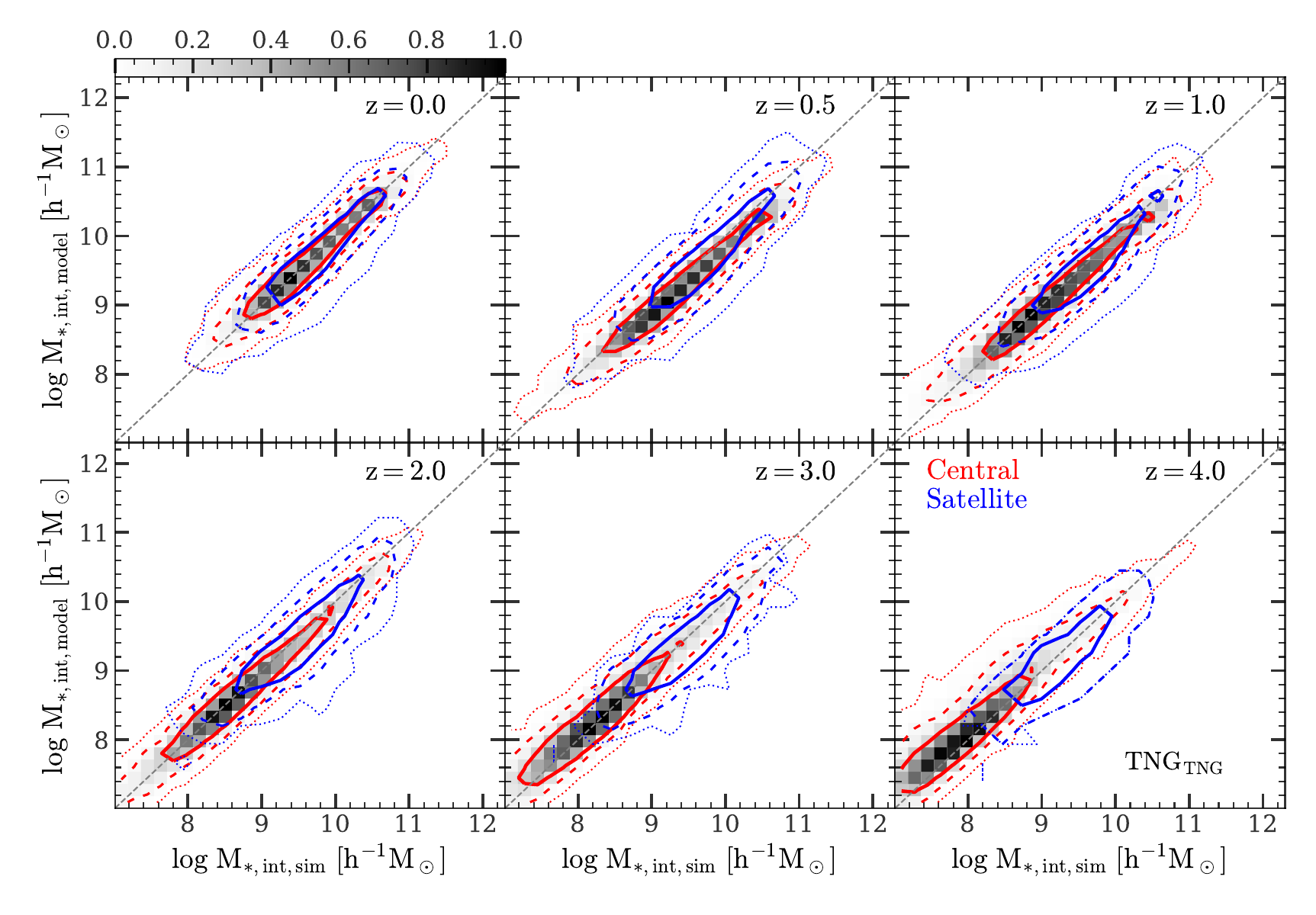}
	\caption{A comparison of galaxy $\MstarInt$ in the model $\rm TNG_{TNG}$ with 
		the TNG simulation. Different panels show the results at different redshifts. 
		In each panel, 
		\textbf{Red} (\textbf{blue}) contours enclose the 1, 2, 3-$\sigma$ regions of 
		central (satellite) 
		galaxies, and
		\textbf{gray shades} are normalized histograms for central galaxies 
		encoded by the color bar.
	}
	\label{fig:ms-ms}
\end{figure*}

\begin{figure*} \centering
	\includegraphics[width=13.5cm]{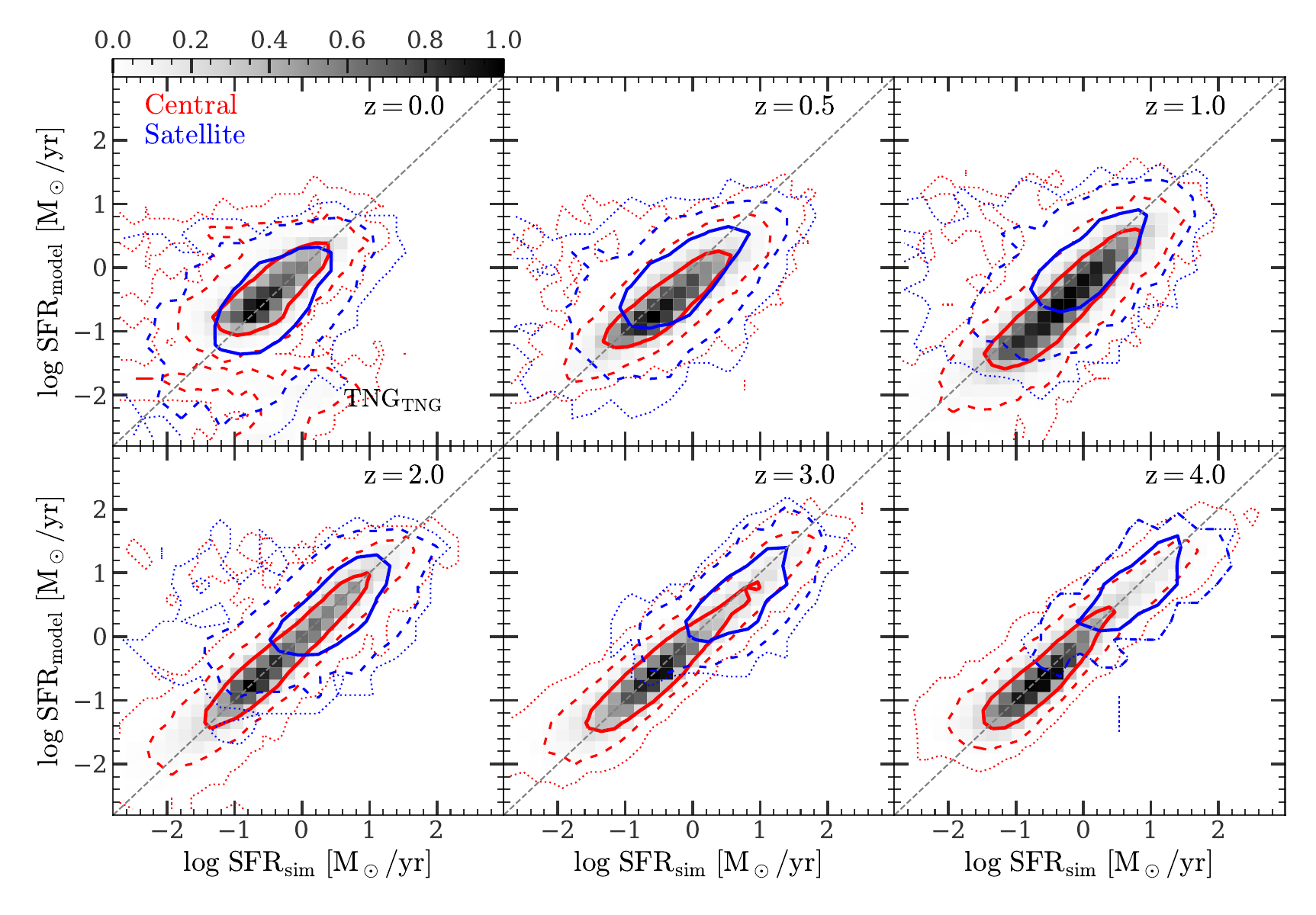}
	\caption{A comparison of galaxy SFR in the model $\rm TNG_{TNG}$ with 
		the TNG simulation. Different panels show the results at different redshifts. 
		In each panel, 
		\textbf{Red} (\textbf{blue}) contours enclose the 1, 2, 3-$\sigma$ 
		regions of central (satellite) 
		galaxies, and 
		\textbf{gray shades} are normalized histograms for central galaxies 
		encoded by the color bar.
	}
	\label{fig:sfr-sfr}
\end{figure*}

\section{Results}
\label{sec:results}

As discussed in \S\ref{sec:intro} and \S\ref{sec:data},  \specialname[MAHGIC] 
can be trained by a hydrodynamic simulation, and 
applied to DMO simulations to make copies.   
Here we use subhalos and merger trees from TNG or EAGLE as 
the training data sets. As a consistency check, the trained model is 
first applied to the hydrodynamic simulation itself, but with all 
the information about the baryonic components discarded.
Because of the impact of baryonic process, dark matter halo properties 
in a hydrodynamic simulation are not expected to be identical to 
those in the corresponding DMO simulation. The check serves as a test 
of the importance of this impact. The model is then applied to the two 
DMO simulations, TNG-Dark and ELUCID. In this section, we show 
results only for the model trained by TNG, while the results
of the EAGLE-trained model are presented in Appendix~\ref{app:eagle-result}.
For convenience, we refer the applications 
of the TNG-trained model to TNG, TNG-Dark, and ELUCID  
as $\rm TNG_{TNG}$, $\rm TNG \smalldash Dark_{TNG}$, and $\rm ELUCID_{TNG}$, 
respectively, while using a subscript `EAGLE' to denote the applications 
of the EAGLE-trained model. 

To achieve a sufficiently high accuracy, we use $N_{\rm piece}=4$ and 
$(z_1, z_2, z_3, z_4)=(0.25, 0.75, 2.5, 5.5)$ to model central galaxies, 
and $N_{\rm piece}=3$ and $(z_1, z_2, z_3)=(0.3, 1.5, 5.5)$ for satellites. 
We note that these choices are made for the training data used here, 
and that different choices can be made when required by the constraining data. 
Tree branches with $\zinfall > 5.5$ are not included in the model, but 
the galaxies in the modeled branches can extend to $z > 5.5$.

\subsection{The importance of individual predictor variables}
\label{ssec:var-contrib}

As discussed in \S\ref{ssec:overall-design}, \specialname[MAHGIC] 
is made interpretable by using the \specialname[PCA] to reduce the 
dimensionality of variables, and the \specialname[GBDT] to build the 
mapping between halos and galaxies. Such interpretability 
enables us to quantify the importance of halo properties to 
a given galaxy property, as well as to estimate the uncertainty 
in the predicted galaxy property.
To measure model uncertainties and the importance of predictor variables, 
we show in Figure~\ref{fig:var-contrib-cent} the cumulative $R^2$ of 
the mapping $\mathbb{R}$ for the model of central galaxies 
trained by TNG. Briefly, $R^2$ is a value between 0 and 1, 
with $R^2=1$ indicating  no uncertainty in the prediction of the target 
variable and $R^2=0$ indicating no correlation between the predictor 
variable and the target variable 
\citep[see,][for a detailed description]{chenHowEmpiricallyModel2021}. 
The $R^2$-value for each target variable is computed by building
a series of \specialname[GBDT] regressors that use 
an increasing number of predictor variables in $\tilde{\bfrm[x]}_{\rm h}$.  
For each regressor, a fraction of $75\%$ of all the branches in the sample 
are drawn randomly without replacement and used as the training set, 
and the remaining $25\%$ are used as the test set to compute $R^2$. 
The training and test processes are repeated 20 times, and 
the standard deviation of the $R^2$-value among them is used as 
an estimate of the error bars. The results shown 
in Figure~\ref{fig:var-contrib-cent} indicate that 
the $\MstarInt$ at the anchor redshift is dominated by the halo mass 
at this redshift, and that more than $80\%$ of the $R^2$ can be achieved 
by using only one halo property. Adding PCs only leads to limited improvements.
This is consistent with the result in Paper-I where it was found 
that the stellar mass and SFR at a given redshift are almost 
totally determined by $v_{\rm max}$, and that adding PCs 
only leads to small variances around the predicted mean SFH.
The $R^2$ value for the target $\rm PC_{*,1}$ is lower, 
typically less than $50\%$, even if all the three 
predictor variables of halos are used. 
For all redshifts, the contribution to $\rm PC_{*,1}$ 
made by the PCs of the subhalo MAH is significant 
in comparison to that of the halo mass,
indicating that the shape of the SFH of a galaxy   
is affected by the shape of the MAH of its host halo. 
The $R^2$-value for $\rm PC_{*,2}$ is small, 
indicating that the detailed variations in the SFH 
are generated by complicated physical processes 
not well captured by using a small set of halo predictors. 
The poor performance of the model on $\rm PC_{*,2}$ 
also indicates that it is not useful to include more 
higher order PCs in the model.

Figure~\ref{fig:var-contrib-sat} shows the cumulative $R^2$ curves of 
the mapping $\mathbb{R}_{\rm infall}$ for the satellite model 
trained by TNG. The $R^2$ curves and error bars are computed 
in the same way as for the central model. Using $M_{\rm h,infall}$, 
$M_{\rm h,infall,cent}$ and $j_{\rm infall}$ is sufficient 
to correctly predict both $I_{\rm merge}$ and $\tau_{\rm merge}$,
where $M_{\rm h,infall,cent}$ is not significant for $I_{\rm merge}$ and 
$j_{\rm infall}$ only helps for $\tau_{\rm merge}$ at high z.
Including $\zinfall$ makes a significant improvement only for 
$\tau_{\rm merge}$ for subhalos in the piece of lowest $\zinfall$. 
However, since only a small fraction ($\sim 3\%$) of the galaxies hosted by 
such subhalos will merge with their centrals by $z=0$, the accuracy of the  
prediction for $\tau_{\rm merge}$ in this piece does not matter much. 
The $R^2$ values that can be reached for $\rm PC_{*,infall,1}$ and $\rm PC_{*,infall,2}$ are 
less than $50\%$ even when all the four halo properties are used, 
indicating that the SFH of a galaxy after infall is affected by 
many nuanced factors and cannot be modeled fully 
by using only a small set of halo properties 
at the infall time. The small $R^2$ for $\rm PC_{*,infall,2}$ suggests 
that including more PCs of the SFH in the model is not helpful. 
The contribution to $\rm PC_{*,infall,1}$ from $\zinfall$ 
is not significant, indicating that the dominant
mode of the SFH after infall does not change significantly over 
each of the redshift intervals in question. 

\subsection{Stellar Mass and Star Formation Rate}
\label{ssec:mstar-and-sfr}

Because the model is trained using TNG, a comparison of 
the predicted galaxy population with that given 
by TNG provides a direct check on the 
the performance of our method. 
Figure~\ref{fig:ms-ms} shows a
comparison of $\MstarInt$ predicted by 
$\rm TNG_{TNG}$ with that given by the TNG simulation for individual galaxies. 
Results for central and satellite galaxies are shown separately at different redshifts.  
As one can see, the model prediction matches
well with the TNG simulation, without any significant
bias. This demonstrates that our multi-stage, non-linear model is flexible enough to capture 
the main properties in the underlying halo-galaxy mapping in the simulation. The relation between
the modeled $\MstarInt$ and the simulated $\MstarInt$ is also tight, with a standard deviation 
typically of 0.14 dex (0.17 dex) for central (satellite) galaxies at $z=0$ 
and 0.23 dex (0.26 dex) at $z=4$. This can be attributed to the use of a piece-wise
approach, in which the mapping is trained over
the whole redshift range. 

Figure~\ref{fig:sfr-sfr} shows the comparison of
$\rm TNG_{TNG}$ with the TNG simulation for 
the SFR of individual galaxies at different redshifts. 
Again, we do not see any significant bias in the 
predictions of $\rm TNG_{TNG}$.
Compared to $\MstarInt$, the relation of the 
SFR between $\rm TNG_{TNG}$ and the TNG simulation
has larger scatter, with a standard deviation 
of 1.2 dex (1.5 dex) for central (satellite) galaxies at $z=0$ and 
0.31 dex (0.46 dex) at $z=4$. This is expected, because SFR 
is a differential property, while $\MstarInt$ is cumulative. As demonstrated in Paper-I, SFR may 
be more sensitive to nuanced factors that are 
difficult to capture using a well-defined set of halo properties.

\begin{figure*} \centering
	\includegraphics[width=16.5cm]{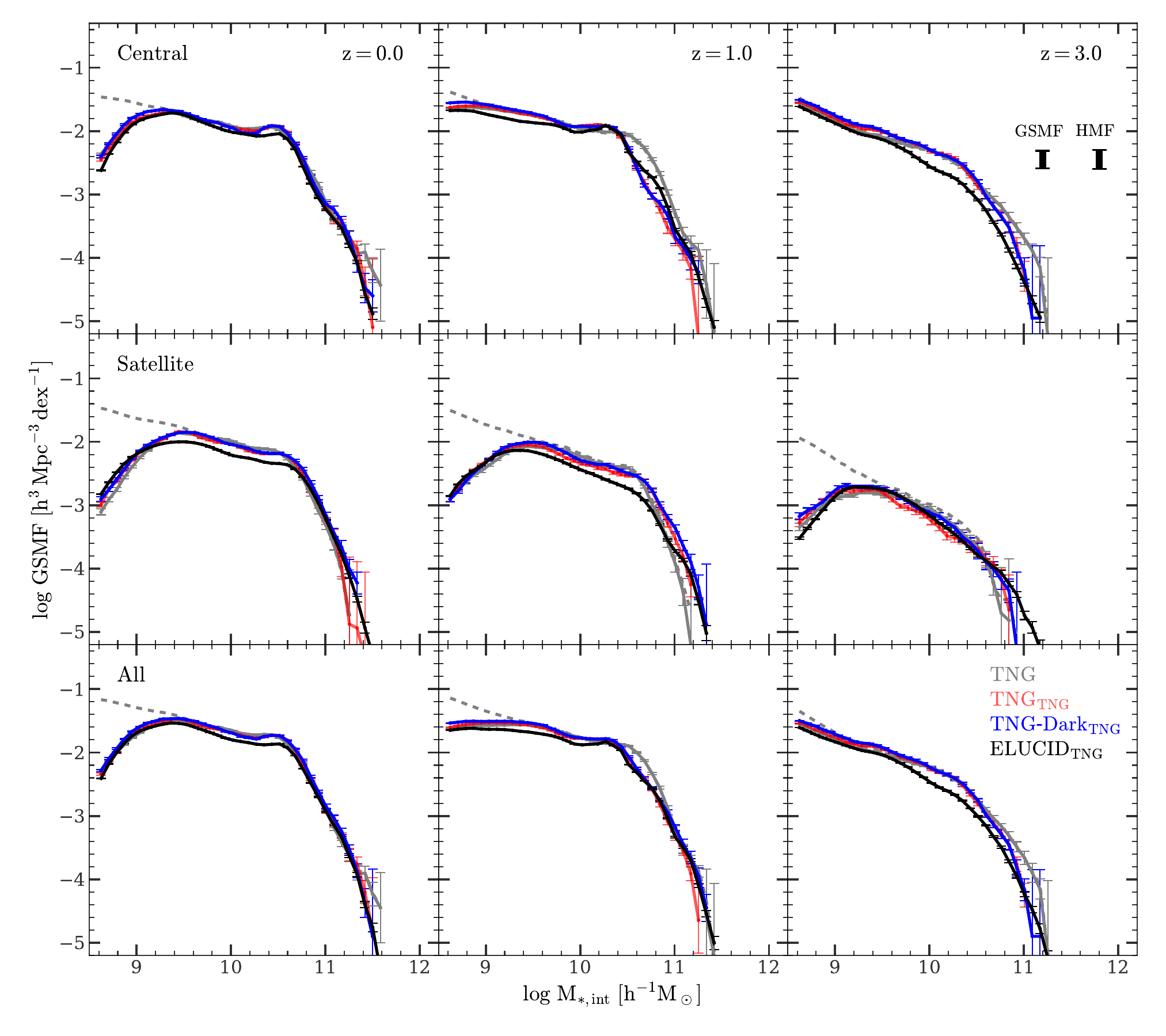}
	\caption{
		GSMFs of modeled galaxies compared with simulated ones. 
		The model is trained using the TNG simulation.
		The \textbf{first}, \textbf{second} and \textbf{third} 
		rows show the GSMFs of central, satellite 
		and all (central+satellite) galaxies, respectively. 
		\textbf{Different columns} show the GSMFs at different redshifts. 
		\textbf{Solid gray} lines are 
		from TNG simulation. \textbf{Red}, \textbf{blue} and \textbf{gray} 
		lines are the results when 
		the model is applied to TNG, TNG-Dark and ELUCID simulations, respectively. 
		\textbf{Dashed gray} lines are GSMFs of all TNG simulated galaxies including 
		those in the subhalos not selected in our samples. 
		Error bars are computed 
		by using 50 bootstrap resamplings. 
		{\revisestyle
		The \textbf{left} of the two vertical bars in the upper right panel 
		indicates the difference in GSMF between
		TNG-$\rm Dark_{TNG}$ and $\rm ELUCID_{TNG}$ at $\MstarInt 
		\in [1.0,5.0]\times 10^{10}\msun$. 
		The \textbf{right} one indicates the difference 
		caused by the difference in HMF between the two simulations.} }
	\label{fig:gsmf}
\end{figure*}

\subsection{Galaxy Stellar Mass Function} 

When the model is applied to a DMO simulation, 
the results cannot be checked on the basis of individual galaxies, 
but can be tested statistically. One of the most important statistical 
properties of the galaxy population 
is the galaxy stellar mass function (GSMF), 
defined as the number density of galaxies as
a function of stellar mass.
Figure~\ref{fig:gsmf} compares the GSMFs obtained 
from TNG, $\rm TNG_{TNG}$, $\rm TNG \smalldash Dark _{TNG}$ and $\rm ELUCID_{TNG}$, separately 
for central, satellite and all galaxies at five different redshifts.
Because our sample selection criteria are based on dark matter halos above a certain mass, as 
described in \S\ref{ssec:samples}, some galaxies are missed in our model. The difference 
between the gray solid curve and the gray dashed line
in each panel of Figure~\ref{fig:gsmf} is caused
by the sample selection. However, comparisons 
can still be made among TNG, $\rm TNG_{TNG}$, $\rm TNG \smalldash Dark _{TNG}$ 
and $\rm ELUCID_{TNG}$
over the entire mass range because the same 
sample selection criteria are used for all of them.

The difference of GSMFs between TNG and $\rm TNG_{TNG}$ is small for both 
central and satellite galaxies at all redshifts. Key features of the GSMF, such 
as the power-law shape at the low-mass end 
and the rapid drop at the high-mass 
end, are well reproduced in $\rm TNG_{TNG}$. 
This is consistent with the result discussed in 
\S\ref{ssec:mstar-and-sfr} that our model has no significant bias 
in $\MstarInt$ and SFR.

The GSMFs obtained from $\rm TNG \smalldash Dark_{TNG} $ are very 
similar to those from $\rm TNG_{TNG}$, for both central and satellite galaxies
and at all redshifts. This is a direct consequence of our model 
strategy (\S\ref{ssec:overall-design}): we intentionally avoid the use of 
predictors that can be significantly affected by baryonic processes. 
In addition, our use of \specialname[PCA] and \specialname[GBDT] 
suppress the complexity of the model so as to avoid over-fitting the 
training data.

In the application of the model to ELUCID, more factors can affect the 
output. From Figure~\ref{fig:gsmf} one can see some noticeable 
differences between $\rm ELUCID_{TNG}$ and $\rm TNG \smalldash Dark_{TNG}$.
For central galaxies at $z \geqslant 1$, the GSMF of $\rm ELUCID_{TNG}$ 
is lower than that of $\rm TNG \smalldash Dark_{TNG}$ in the intermediate mass range. 
This is produced, at least partly, by the lower halo mass function (HMF)
of ELUCID at $z \geqslant 1$ compared to TNG-Dark,
as shown in Appendix~\ref{app:subhalo-history}. It may also 
be produced by a difference in the MAH of halos between the two 
simulations, although this difference is quite small, 
as shown in Appendix~\ref{app:subhalo-history}.
To see the effect of the lower HMF, 
we compute the difference of the GSMF between $\rm ELUCID_{TNG}$ and $\rm TNG \smalldash Dark_{TNG}$, 
and compare it with the difference in HMF. The comparison is shown in
the top right panel of Figure~\ref{fig:gsmf} using
{\revisestyle two bars} for central galaxies with 
$\MstarInt \in [1.0, 5.0] \times 10^{10} \msun$ at {\revisestyle $z=3$, 
where the difference in GSMF is the most significant. 
The difference in the GSMF is 0.28 dex, while the difference in the 
HMF is 0.30 dex in the corresponding halo mass range 
estimated from the $\MstarInt$-$\Mhalo$ relation. From this we 
conclude that the difference in the HMF, produced by the cosmic variance, 
is the most significant contributor to the difference in the GSMF.}
For satellites, the difference between 
$\rm ELUCID_{TNG}$ and $\rm TNG \smalldash Dark_{TNG}$ is small at high-$z$, but 
becomes larger at $z\leqslant 1$, where the GSMF 
of ELUCID is lower than that of $\rm TNG \smalldash Dark_{TNG}$ in the intermediate 
mass range. This is expected, given that $\rm ELUCID_{TNG}$ underestimates 
the number of central galaxies at $z>1$, which are potentially 
the progenitors of the satellite galaxies at lower $z$. 

The GSMFs of the total population (centrals and satellites together) are
shown in the third row of Figure~\ref{fig:gsmf}. We see again 
that $\rm TNG_{TNG}$ and $\rm TNG \smalldash Dark _{TNG} $ are both 
in good agreement with the TNG simulation. $\rm ELUCID_{TNG}$ 
underestimates the GSMF at $z>1$ in the intermediate stellar mass range.
In general, the GSMF is dominated by centrals, more so at higher $z$. 

\begin{figure*} \centering
	\includegraphics[width=17cm]{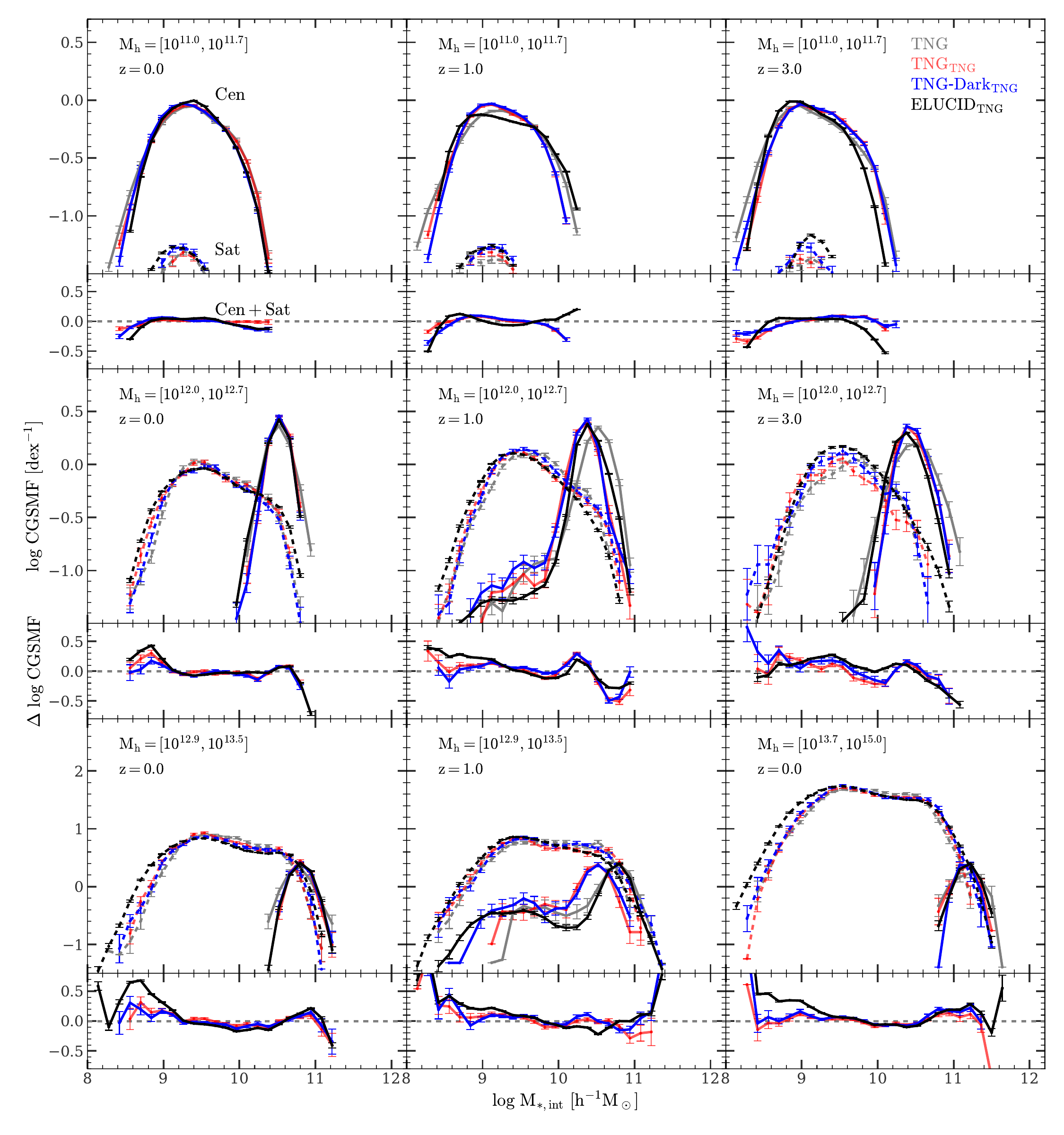}
	\caption{CGSMFs of dark matter halos with different halo masses $\Mhalo / (\msun)$ 
		and at different redshifts as indicated in each panel. In each panel,
		\textbf{gray} lines are 
		from the TNG simulation. \textbf{Red}, \textbf{blue} and \textbf{black} 
		lines are results when 
		the model is applied to TNG, TNG-Dark and ELUCID, respectively.  
		\textbf{Solid} and \textbf{dashed}
		lines are for central and satellite galaxies, respectively.
		Error bars are computed by using 50 bootstrap resamplings. 
		{\revisestyle 
		Each small panel shows the model residuals of CGMSFs
		relative to the TNG simulation for all (central+satellite) galaxies. }
		}
	\label{fig:cgsmf}
\end{figure*}

\subsection{Conditional Galaxy Stellar Mass Functions}
\label{ssec:cgsmf}

To check our model in more detail, we examine the conditional galaxy stellar mass 
function (CGSMF) for halos of different mass. This is a cleaner 
test, as it is not affected by variations in the halo mass  
function between the training simulation (TNG) and the target
simulation (ELUCID) due to cosmic variance. 
Figure~\ref{fig:cgsmf} shows the CGSMFs of centrals and satellites 
in FoF halos of different masses at different redshifts. 
Note that we do not show results for high-mass halos at high $z$
because such halos are too rare to give a reliable CGSMF.    
The results obtained from $\rm TNG_{TNG}$, $\rm TNG \smalldash Dark _{TNG}$ 
and $\rm ELUCID_{TNG}$ are shown together and compared to   
the corresponding results obtained directly from the TNG simulation.
As one can see, the CGMSFs of low-mass halos are dominated by 
centrals at all redshifts. As the halo mass increases, 
the peak of the central CGSMF moves rightward, as a result of the 
halo mass - central stellar mass relation. 
For high-mass halos, satellites dominate the CGSMF. 
For high-mass halos at high-$z$, for example, {\revisestyle those with 
$\Mhalo / (\msun) \in [10^{12.9}, 10^{13.5}]$ 
at $z=1$}, the CGMSFs of centrals in 
the TNG simulation have a tail at the low-mass end.  
Our model captures this feature, although it is unclear 
if the feature is physical.

In terms of the CGSMF, we see that  
$\rm TNG_{TNG}$ and $\rm TNG \smalldash Dark _{TNG}$
match closely with each other at all redshift and for halos 
of different mass, and that both are compatible with the TNG simulation. 
This demonstrates again that our model is not significantly affected 
by uncertainties introduced by baryonic processes. 
The CGSMFs obtained from $\rm ELUCID_{TNG}$ also 
closely follow the TNG-based results, although some 
differences are noticeable. In general, the differences 
are not much larger than the variances among the TNG-based results, 
indicating that our model is valid for any DMO simulation 
where the dark matter halo population is modeled reliably.   

\begin{figure*} \centering
	\includegraphics[width=16cm]{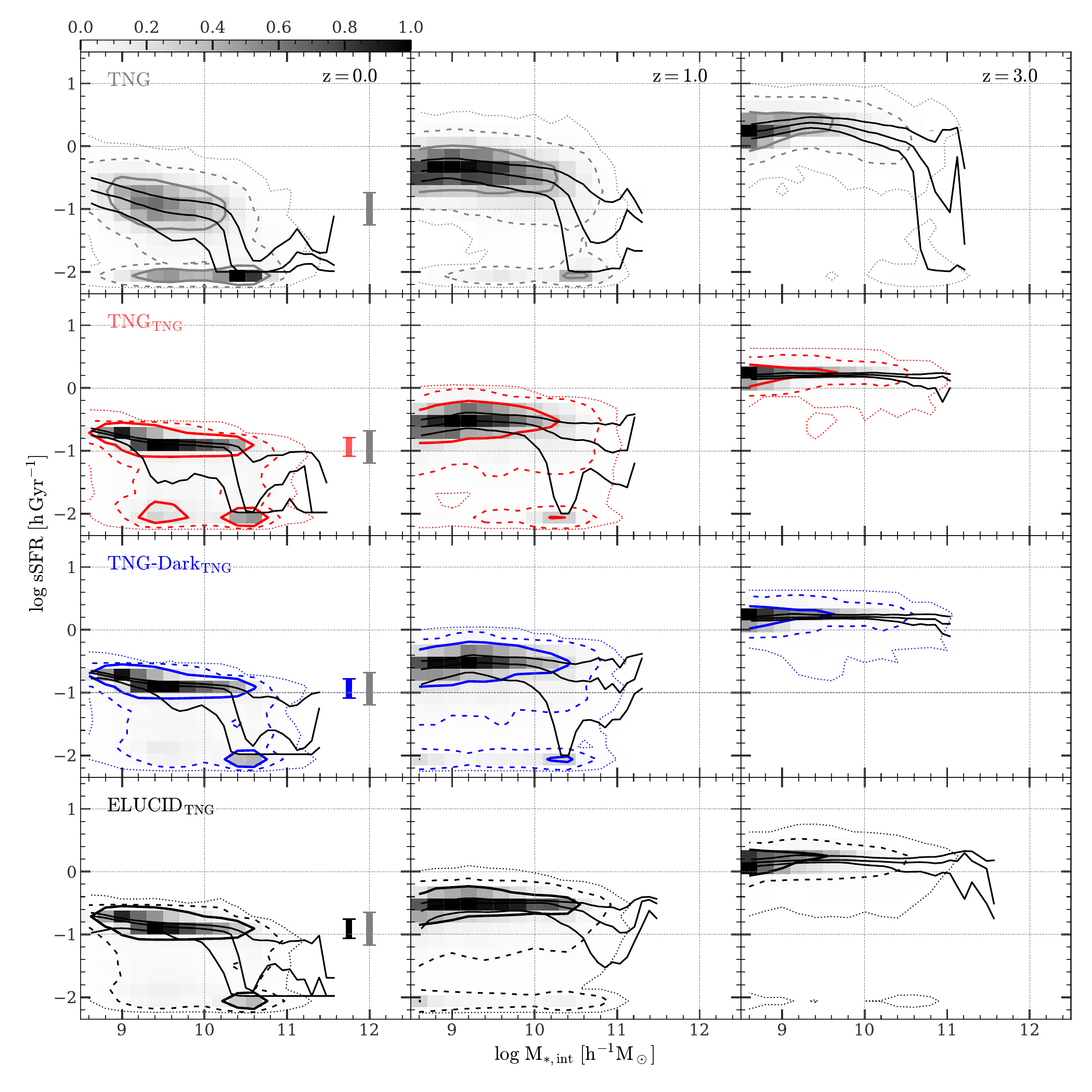}
	\caption{The galaxy distribution in the $\log \MstarInt$-$\log$ sSFR plane. 
	Different \textbf{columns} show the results at different redshifts indicated in the 
	first row. The \textbf{first row} shows the result from the TNG simulation. The \textbf{other 
	three rows} show the results when the model is applied to TNG, TNG-Dark and 
	ELUCID, respectively. 
	In each panel, \textbf{gray} shades are normalized histograms encoded by 
	the color bar. \textbf{Solid}, \textbf{dashed} and \textbf{dotted} 
	contours enclose 1,2,3-$\sigma$ regions,
	respectively. The \textbf{black solid} lines show 
	the $25\%$, $50\%$ and $75\%$ quantiles 
	of $\log {\rm sSFR}$ at a given $\MstarInt$. 
	{\revisestyle
	The \textbf{gray bar} in the top left panel 
	indicates the 1-$\sigma$ scatter of the main sequence at 
	$\MstarInt \sim 10^{10}\msun$ 
	computed directly from the simulated data. 
	The \textbf{colored bars} in other left panels show 
	the scatter of the modeled galaxies while the \textbf{gray bars} 
	show the scatter after the missed random component is added.
	}}
	\label{fig:bimodality}
\end{figure*}

\subsection{The Star-forming Main Sequence and Galaxy Bimodality}
\label{ssec:bimodality}

The galaxy distribution in the $\rm \log sSFR$ - $\log \MstarInt$ space 
is observed to be bimodal. The mode with high sSFR is referred to as 
the star-forming main sequence, while the one with lower sSFR 
is referred to as the quenched population. This bimodal distribution 
was first established observationally \citep[e.g.][]{
	stratevaColorSeparationGalaxy2001,
	blantonBroadbandOpticalProperties2003,
	baldryQuantifyingBimodalColorMagnitude2004,
	liDependenceClusteringGalaxy2006,
	faberGalaxyLuminosityFunctions2007,
	brammerDEADSEQUENCECLEAR2009,
	coilPRIMUSMathplusDEEP22017}, 
and later reproduced  in some hydrodynamic simulations, such as 
the TNG simulation \citep{nelsonFirstResultsIllustrisTNG2018}. 
The bimodal distribution 
contains important information about galaxy formation and evolution, 
and should be reproduced in any successful model. 

Figure~\ref{fig:bimodality} shows, in the first row, the distribution of galaxies 
at different redshifts in the $\rm \log sSFR \smalldash \log \MstarInt$ plane 
obtained directly from the TNG simulation. The predictions of 
$\rm TNG_{TNG}$, $\rm TNG \smalldash Dark _{TNG}$ and $\rm ELUCID_{TNG}$
are shown in the subsequent rows, respectively. 
Because of the limited resolution, TNG cannot resolve the
star formation activity reliably when the SFR is too low. 
These low-SFR galaxies are stacked at the bottom 
of each panel in the first row of Figure~\ref{fig:bimodality}, 
which has the effect of artificially making the 1-$\sigma$ contours 
of the quenched population tight. As one can see, TNG galaxies 
show a strong bimodal distribution at low $z$, but the 
quenched population decreases with increasing $z$ and 
{\revisestyle disappears at $z = 3$}. The majority of the galaxy population 
in the TNG simulation start from a well-defined star-forming 
sequence at high $z$, and become quenched subsequently. 
The quenching starts from the massive end at $z\sim 3$ and moves to 
lower mass at lower $z$. 

The results obtained from the three applications, 
$\rm TNG_{TNG}$, $\rm TNG \smalldash Dark _{TNG}$ and $\rm ELUCID_{TNG}$, 
are all comparable with each other. At a given $z$, all the models predict 
star-forming main-sequences with a similar amplitude and dispersion. 
The predicted sequences have amplitudes similar to those 
given by the TNG simulation, but with smaller dispersion, 
particularly at $z=0$. This indicates that our model, which is based on a
limited number of predictors (halo properties), is not able to 
capture all the variances in the SFR. As demonstrated in Paper-I, 
more than $50\%$ of the main sequence scatter at $z=0$, 
as measured by the $R^2$ of the \specialname[GBDT] regressors, 
is contributed by nuanced factors that are difficult to relate 
to halo properties. One possible way to correct for this, as 
proposed in Paper-I, is to include a random component, 
whose variance is characterized by the 
difference between the required variance and the 
modeled variance of $\log {\rm sSFR}$ in the main sequence.
Here we can obtain this by comparing the main sequence galaxies 
in the simulation and in our model. {\revisestyle The vertical bars
in} Figure~\ref{fig:bimodality} show the effects of adding a random 
component to the modeled galaxies at $z=0$. The main-sequence scatter is enlarged 
significantly for the three application cases, and the predicted scatter
is now comparable to that given by the TNG simulation. Overall, 
the predicted 2-$\sigma$ contours in the three application cases are 
comparable to those obtained from the TNG simulation.
At low $z$, our model predicts an extended quenched population, consistent with 
the simulation, but the predicted  quenched population 
at the high-stellar-mass end is smaller than in the 
TNG simulation, as seen from the $50\%$ and $75\%$ percentile lines. 
This discrepancy arises from the difficulty in predicting whether or not 
a high-mass galaxy is quenched solely on the basis of 
halo properties, as shown in Paper-I. The predicted quenched 
population in the low stellar mass range at $z=0$ is also more 
diffuse in the three application cases. As shown in Paper-I, 
this is a result of sample imbalance: low-mass 
galaxies in TNG at $z \sim 0$ are mainly star-forming galaxies, 
and our model is more concentrated on the star-forming population, 
leaving the quenched population less well modeled.

\begin{figure*} \centering
	\includegraphics[width=18cm]{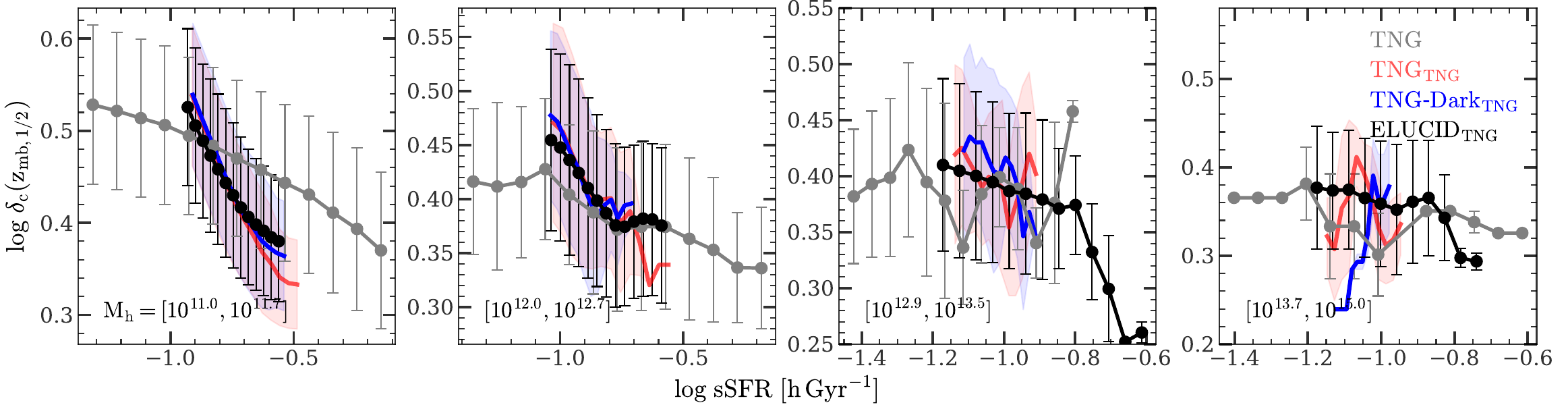}
	\caption{
		The relation between subhalo formation time $z_{\rm mb,1/2}$ and 
		sSFR for $z=0$ central subhalos. Only the star-forming population 
		(${\rm sSFR} \geqslant 10^{-2} \gyr$) is shown. Each panel shows the 
		result for subhalos with a given halo mass as indicated at the lower left corner 
		of the panel. \textbf{Gray} lines show the result from TNG simulation. 
		\textbf{Red}, \textbf{blue} and \textbf{black} lines are results 
		when we apply the TNG-trained model to TNG, TNG-Dark and ELUCID, 
		respectively. Error bars and shades are standard deviations in 
		bins. }
	\label{fig:assembly-bias}
\end{figure*}

\subsection{The Correlation of Star Formation with Halo Assembly}
\label{ssec:assembly-bias}

Because of the inclusion of MAH PCs in our model, galaxy properties 
predicted by the model are naturally correlated with the MAH of the host 
halos. In Figure~\ref{fig:assembly-bias}, we show 
the relation between the halo half-mass formation time, 
$z_{\rm mb,1/2}$, and the current sSFR for central 
star-forming galaxies at $z=0$. 
Here $z_{\rm mb,1/2}$ is calculated by tracing the main branch in a 
subhalo merger tree rooted in the target subhalo, and the redshift  
is represented by  
\beq
\delta_{\rm c}(z)=\frac{\delta_{\rm c,0}}{D(z)},
\eeq
where $\delta_{\rm c, 0}=1.686$ is the critical overdensity given 
by the spherical collapse model, and $D(z)$ is the linear growth 
factor at $z$ given by \cite{carrollCosmologicalConstant1992}. 

As described at the end of \S\ref{sec:model} and shown in \S\ref{ssec:bimodality},
the modeled sSFR is missing a random component that cannot be fully explained
by the halo properties considered here. Consequently, the 
predicted sSFR for star-forming galaxies spans a smaller range than 
that given by the simulation. The relatively small dynamic 
range in $\log {\rm sSFR}$ shown in Figure~\ref{fig:assembly-bias} for 
the three application cases is caused by this. 
Taking into account the random component, our model actually 
reproduces the trends seen in the simulation: galaxies in halos of  
larger $\delta_{\rm c}(z_{\rm mb,1/2})$ tend to have smaller sSFR. 
The only exception is for massive systems, where the uncertainty 
is too large to see the correct trend in $\rm TNG_{TNG}$ and 
$\rm TNG \smalldash Dark_{TNG}$. We also note that 
the correlation between halo assembly and sSFR is weak, and the 
variance between individual galaxies is large, as can be seen from the 
large error bars and shadings shown in Figure~\ref{fig:assembly-bias}.

\begin{figure*} \centering
	\includegraphics[width=18cm]{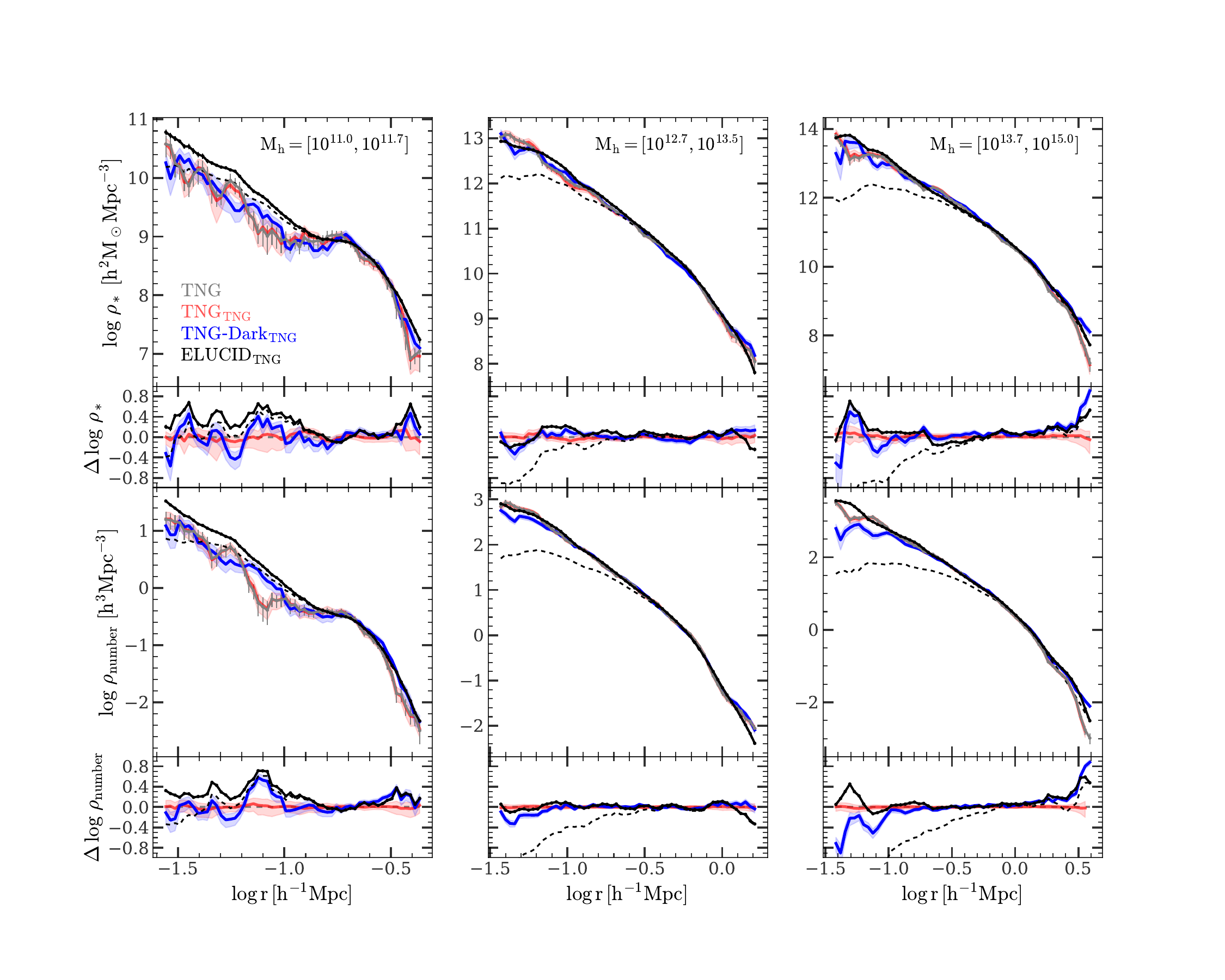}
	\caption{ The $z=0$ $\MstarInt$ density profile (\textbf{first row}) 
		and number density profile (\textbf{second row})
		of satellite galaxies in dark matter halos with different $\Mhalo/(\msun)$ 
		indicated in the panels of first row. 
		In each panel, \textbf{gray} 
		line is from TNG simulation. \textbf{Red}, \textbf{blue} and 
		\textbf{solid black} lines are results when the model is applied to 
		TNG, TNG-Dark and ELUCID respectively. 
		The \textbf{dashed black} line is also for ELUCID, but only 
		shows the subhalos that are resolved by the simulation.
		Error bars and shades represent the standard deviation 
		computed using 50 bootstrap resamplings. 
		{\revisestyle Each small panel shows the residuals of curves relative to 
		the gray line in the same panel.}
		}
	\label{fig:density-profile}
\end{figure*}

\begin{figure*} \centering
	\includegraphics[width=16.5cm]{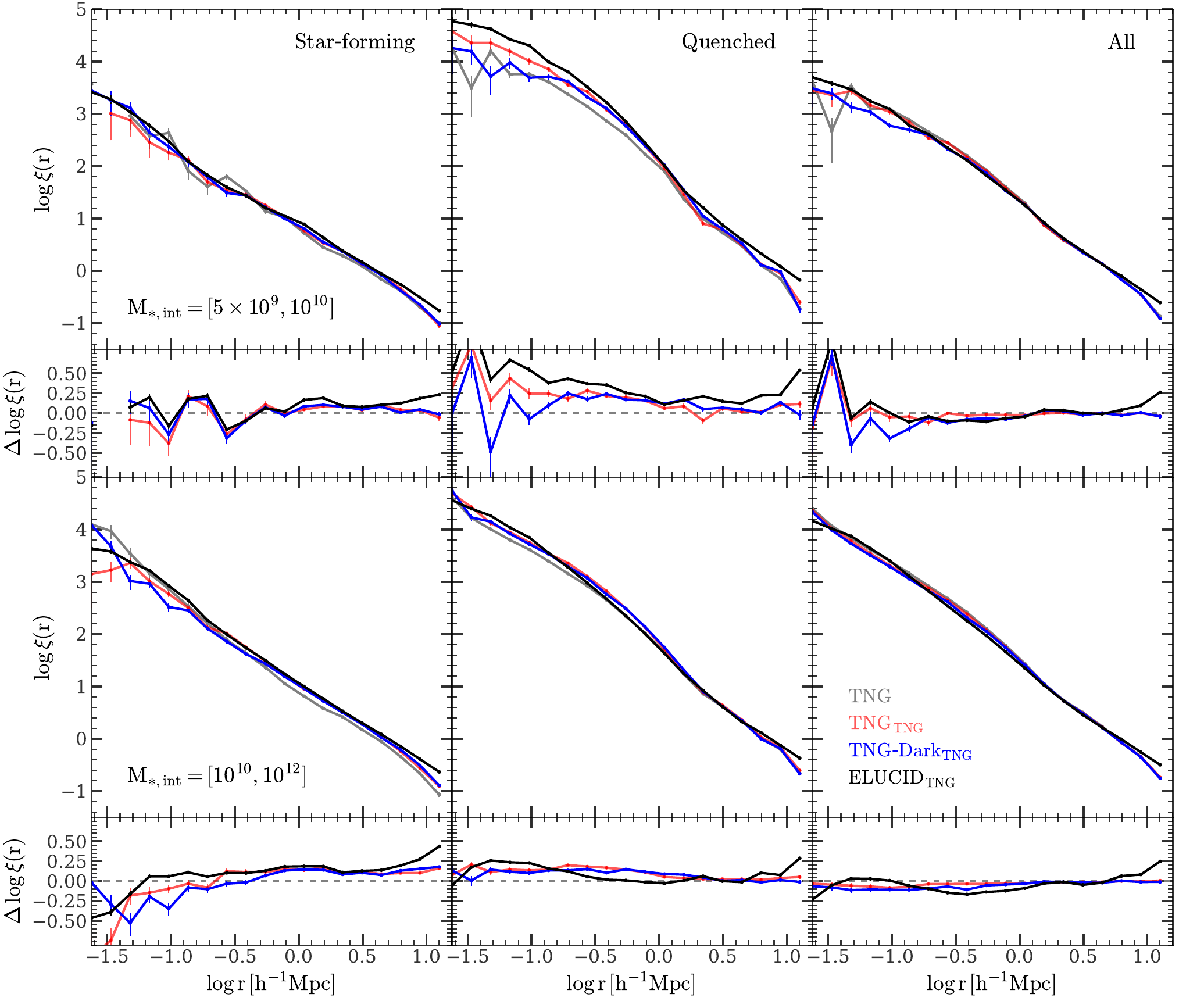}
	\caption{ \revisestyle
		The auto-correlation functions for galaxies at $z=0$
		with different $\MstarInt / (\msun)$ ({\bf rows}). 
		Results for star-forming, quenched, 
		and all (star-forming + quenched) galaxies
		(divided at ${\rm sSFR} = 5\times 10^{-1} \gyri$) are shown separately 
		in the three columns.
		Error bars represent the standard deviation 
		computed using 50 bootstrap resamplings. 
		In each panel, \textbf{gray} 
		line is from TNG simulation. \textbf{Red}, \textbf{blue} and 
		\textbf{black} lines are results of the model applied to 
		TNG, TNG-Dark and ELUCID, respectively. 
		Each small panel shows the residuals of the corresponding curves 
		relative to the gray line in the same panel. }
	\label{fig:corr-func}
\end{figure*}

\subsection{The Spatial distribution of satellite galaxies in dark matter halos
and the two-point correlation function}
\label{ssec:density-profile}

Galaxy clustering is a commonly used statistical property to characterize the 
spatial distribution of the galaxy population.
From a theoretical perspective, the galaxy-galaxy correlation can be decomposed 
into two components: the `two-halo term', which describes the correlation 
produced by the halo-halo correlation, and the `one-halo term',
which is produced by the galaxy distribution in individual dark matter halos. 
The two-halo term on large scales is determined as long as the halo occupation 
of galaxies is correctly modeled.
The `one-halo term', on the other hand, depends on the details of the galaxy 
distribution in halos. Since our model reproduces the CGSMF in halos of different 
mass (\S\ref{ssec:cgsmf}), 
it is already tested for its predictions for halo occupation. Here
we first present test results for the predicted galaxy distribution inside halos.

In \specialname[MAHGIC], galaxies are modeled based on their host subhalos 
and their positions are assumed to be the same as the positions of the 
most bound particles of the corresponding subhalos. Thus, as long as the DMO 
simulations correctly predict the distribution of subhalos in their host halos, 
the galaxy distribution will also be reproduced as long as the subhalo-galaxy 
interconnection is correctly predicted. 
Figure~\ref{fig:density-profile} shows the stellar mass density profile 
and number density profile of satellite galaxies in halos of different masses 
at $z=0$. The three application cases, $\rm TNG_{TNG}$, 
$\rm TNG \smalldash Dark _{TNG}$ and $\rm ELUCID_{TNG}$ are all plotted and 
compared to the TNG simulation. As described in \S\ref{ssec:model-of-satellite}, 
we have to follow some satellite galaxies in ELUCID using merger times 
calibrated with high resolution-simulations, because their subhalos 
are not well resolved in ELUCID. These galaxies do not have simulated 
subhalos associated with them in ELUCID and, therefore, do not have 
subhalo-based positions. The results obtained without these galaxies   
are shown in Figure~\ref{fig:density-profile} as the dashed black curves. 
We can look at the results in three steps. First, the results between TNG and 
$\rm TNG_{TNG}$ are almost indistinguishable for halos of different mass, 
indicating that the subhalo-galaxy interconnection predicted by our model is
unbiased. This is consistent with the results, presented in \S\ref{ssec:mstar-and-sfr},
that the predicted $\MstarInt$ and SFR follow the simulation results closely.
Second, comparing the results of $\rm TNG_{TNG}$ and $\rm TNG \smalldash Dark_{TNG}$, 
we see that baryonic processes have some effects on the results. 
The red and blue lines in Figure~\ref{fig:density-profile} 
follow each other tightly over a large range of $r$. Some differences 
can be seen in the inner regions of halos ($r<0.1 \mpc$), where the  
number density of satellites predicted by $\rm TNG \smalldash Dark_{TNG}$
is lower than that of $\rm TNG_{TNG}$, {\revisestyle particularly for halos with
$M_{\rm h} \geqslant 10^{12.7} \msun$}.
Apparently, the baryonic component in a satellite can make its 
subhalo more concentrated and harder to destroy by environmental 
effects near the center of its host halo. 
The difference in the stellar mass density profile is smaller than that 
in the number density profile, indicating that the destroyed 
subhalos are preferentially of low mass. 
These are consistent with the results obtained by \cite{simhaTestingSubhaloAbundance2012}
using subhalo abundance matching technique, where they found that 
the stellar mass loss can significantly affect the radial profile of 
low-mass galaxies in massive halos.
Third, comparing the results 
between $\rm TNG \smalldash Dark_{TNG}$ and $\rm ELUCID_{TNG}$, 
we can clearly see the effects of the simulation resolution. 
The difference between the blue and dashed black curves in 
Figure~\ref{fig:density-profile} 
is significant for halos above $M_{\rm h}=10^{12} \msun$.
This is expected, because the ELUCID simulation has a much lower resolution and
subhalos can go below the mass resolution limit or be disrupted artificially  
before they merge with the central subhalo \citep[e.g.,][]{greenTidalEvolutionDark2021}. 
The underestimates of the galaxy stellar mass and number densities are 
more significant in higher mass halos because, at a given
$r$, the mass density is higher in halos of higher mass. 
At large $r$ where subhalos are resolved in ELUCID, the predicted 
profiles tightly follow those of the TNG simulation. Note
that we only use halo properties at the infall time to 
model satellite galaxies, which has the advantage 
of being independent of resolution issues and artificial disruption 
after infall.

To account for the numerical effects on 
the density profile of satellite galaxies, we match each 
FoF halo in ELUCID to a FoF halo in TNG-Dark that has the same 
redshift and $\Mhalo$. The satellite galaxies in ELUCID that do not have
associated subhalos are assigned positions using the subhalos in the 
matched halo from TNG-Dark. The results obtained by including these 
subhalos are shown in Figure~\ref{fig:density-profile} as the 
solid black curves. The match between the TNG-Dark and the ELUCID results 
are now much better for all halos with $M_{\rm h} \geqslant 10^{12.0} \msun$.
For small halos with $\Mhalo \in [10^{11}, 10^{11.7}] \msun$, the 
comparison cannot be easily made, because the TNG and TNG-Dark results 
have large fluctuations owing to the limited sample size. 
This again demonstrates the effects of cosmic variance
and the importance of combining the two types of simulations
to construct statistically reliable mock samples.

{\revisestyle 
For completeness, Figure~\ref{fig:corr-func} shows  
the auto-correlation function for the total galaxy population, 
and separately for star-forming and 
quenched populations in two stellar mass bins. 
In all cases, there is a moderate 
difference between the TNG simulation and our model results. 
The largest difference is seen for the quenched low-mass population 
dominated by small satellite galaxies. 
This is the case because of the inaccuracy of our model in 
predicting the quenched fraction in the low-mass population, 
as discussed in \S\ref{ssec:bimodality}. 
}

To conclude, our model correctly reproduces the satellite 
distribution in dark matter halos, as long as their subhalos 
can be resolved in the DMO simulation. For subhalos whose 
positions cannot be followed reliably in a large-volume DMO simulation,  
their positions can be modeled statistically using
calibrations based on high-resolution DMO simulations. Thus, 
\specialname[MAHGIC] also provides a reliable prescription to model 
the spatial distribution of galaxies.  

\section{Summary and Discussion}
\label{sec:summary}

In this paper, we develop a model, \specialname[MAHGIC] 
({\it Model Adapter for the Halo-Galaxy Inter-Connection}), 
to establish the interconnection between galaxies and dark matter halos.
The model uses a set of halo (subhalo) properties, 
such as halo mass, MAH and orbit, as model input, and transforms it into 
a set of galaxy properties, such as stellar mass and SFH.
We use PCA and GBDT to help the model design, and incorporate 
them into the model pipeline.
We use two sets of hydrodynamic simulations, 
TNG and EAGLE, to train the model, and apply it to a large DMO simulation, 
ELUCID, to demonstrate the reliability, flexibility and accuracy of our model.
The key points and the main results that we obtain in the feature selection, 
model design, training and testing are summarized below.

We select a set of subhalo properties and a set of galaxy properties 
as the predictors and target variables of the model, respectively.
This \textbf{feature selection}, based on the methods and results 
described in \cite{chenRelatingStructureDark2020} and Paper-I, 
can be summarized as follows:
\benum
\item 
Only the most important subhalo properties 
are selected as the predictors of galaxy stellar properties. 
Properties (e.g., halo structural and environmental properties) 
that are strongly degenerate with other more 
important properties (e.g., halo mass, MAH) are not used. 
We also avoid the use of subhalo properties 
that are sensitive to baryonic processes. Subhalo properties 
that are not well resolved in large DMO simulations, 
such as the halo MAH at high redshift and the surviving time of subhalos
after being accreted by their hosts, are modeled using 
calibrations of high-resolution simulations. The final set of subhalo 
properties used by \specialname[MAHGIC] include halo mass, the MAH and the orbit. 
The set of stellar properties used as the target variables are stellar mass
and SFH of individual galaxies. These quantities can be used to obtain
the SFR and quenching status at any given redshift. 
\item 
We apply PCA to the MAH of subhalos and the SFH of galaxies, and 
transform them into sets of PCs in spaces of lower dimensions. 
This data compression step gives a set of linearly-independent 
PCs, further reducing the degeneracy among subhalo properties.
The use of PCs can reduce the model complexity by eliminating 
high-order PCs that are not constrained by the data. It 
also makes the model adaptable - more PCs can be included 
to accommodate additional constraints from new data, 
as guided by the Bayesian theory. As shown in \citet{chenRelatingStructureDark2020},
PCs of halo MAH are interpretable because of their linear nature, 
as seen from their tight correlations with various quantities 
characterizing the formation of subhalos.
\eenum

\specialname[MAHGIC] provides a full pipeline to map subhalo properties 
to galaxy properties, and the main steps in its construction 
are summarized below.
\benum 
\item 
We use GDBTs to map subhalo properties to galaxy properties in spaces of 
reduced dimensions. The tree-based method is capable for building 
highly non-linear relationships between variables, which is important
for our problem. GBDT uses an ensemble of randomized trees to overcome over-fitting, 
hence ensuring the robustness of our model. GBDT also provides summary statistics,
such as feature importance $\mathcal{I}$ and $R^2$, to help understand the 
interaction among different variables, making the model interpretable.
\item 
We model central and satellite galaxies separately, and break 
the reconstruction of the SFH into several redshift pieces. This multi-component 
and multi-stage treatment of the SFH allows the model to be adapted to 
the availability of constraining data at different redshifts.
\item 
As a demonstration of the performance of \specialname[MAHGIC],
we use the hydrodynamic simulation, TNG, to train the model, and apply the trained 
model to dark matter halos given by TNG, TNG-Dark and ELUCID.
The comparison between the TNG results and the outputs of these 
applications verifies that the model is both reliable and flexible. 
We also train our model using an independent hydrodynamic simulation, EAGLE, 
and the results provide further support to our conclusion.
\item 
The test using DMO simulations shows that our model can reproduce a variety of 
statistical properties of the galaxy populations in the training hydrodynamic 
simulations. The predicted $\MstarInt$ and SFR for individual galaxies are unbiased, 
with a well-controlled dispersion relative to the true values.    
The GSMFs at different redshifts, and the CGSMFs in halos of different $\Mhalo$, 
are well recovered. The star-forming main sequence and the galaxy 
bimodality are well captured by our model.
Even the weak correlation between galaxy sSFR and halo assembly 
time is also reproduced. Finally,  the model prediction 
for the spatial distribution of galaxies in their host 
halos also matches that given by hydrodynamic simulations, 
indicating that \specialname[MAHGIC] is capable of modeling galaxy 
clustering.
\eenum

The reliability, accuracy and flexibility of  \specialname[MAHGIC]
in recovering galaxy statistical quantities indicate that it can be 
used to make mocking copies of hydrodynamic simulations into DMO 
simulations with larger volumes. The copied galaxy population shares 
the same halo-galaxy interconnection with the training 
hydrodynamic simulation, while the larger sample provided by 
the DMO simulation provides a fairer representation of the 
galaxy population expected from the physical processes assumed in 
the hydrodynamic simulation. Therefore, \specialname[MAHGIC] provides 
an adapter to link these two types of simulations and to 
combine their individual advantages. 

The general framework provided by \specialname[MAHGIC] can be extended 
to another type of applications that use observational data as constraints. 
In this case, the PCA templates of SFH trained from hydrodynamic simulations 
can still be used to reduce the complexity of the model, as long as 
the templates reflect the real star formation modes in the universe.
Other parts of the pipeline can be adjusted accordingly to the observational data. 
For example, if the SFH cannot be obtained reliably from 
the observation, the corresponding variables, $\tilde{\bfrm[x]}_*$, in the 
representation layer of galaxies, will become latent variables. 
The regressor, $\mathbb{R}$, can also be adjusted to a differentiable form, 
such as that of polynomials or neural networks, which allows the optimization 
to be made by a gradient-based method, such as the back-propagation algorithm 
\citep[e.g.,][]{rumelhartLearningRepresentationsBackpropagating1986}. 
{\revisestyle Some optimization methods that do not require gradient computation 
may also be useful, such as the particle swarm method used 
by \cite{mosterGalaxyNetConnectingGalaxies2020} in the 
optimization of a network-based model.}
{\revisestyle 
Other optimization methods specific to models with latent variables, 
such as the expectation-maximization (EM) algorithm, the more general 
variational inference algorithm, and the sampling method for Bayesian 
networks \citep[e.g.,][]{bishopPatternRecognitionMachine2006, luBayesianApproachSemianalytic2011, 
luBayesianInferenceGalaxy2012,behrooziUniverseMachineCorrelationGalaxy2019a}
}, may also be used to optimize the model. 
We will come back to this when we apply the model to observational 
data.

\section*{Acknowledgements}

This work is supported by the National Key R\&D Program of China 
(grant No. 2018YFA0404502, 2018YFA0404503), and the National Science Foundation 
of China 
(grant Nos. 11821303, 11973030, 11673015, 11733004, 11761131004, 11761141012, 
11890693).
We acknowledge Dandan Xu, Yuning Zhang and Jingjing Shi in the 
access of TNG simulation data. YC and KW gratefully 
acknowledge the financial support from China Scholarship Council.

\section*{Data availability}
The data and software used in this article will be available  
from the corresponding author. They are also available at
\url{https://lig.astro.tsinghua.edu.cn/}. 

The computation was supported by the HPC toolkit \specialname[HIPP] at 
\url{https://github.com/ChenYangyao/hipp}.

\bibliographystyle{mnras}
\bibliography{references}

\appendix
\section{Robustness of Subhalo Properties in the Simulations}
\label{app:subhalo-history}

\begin{figure} \centering
	\includegraphics[width=\columnwidth]{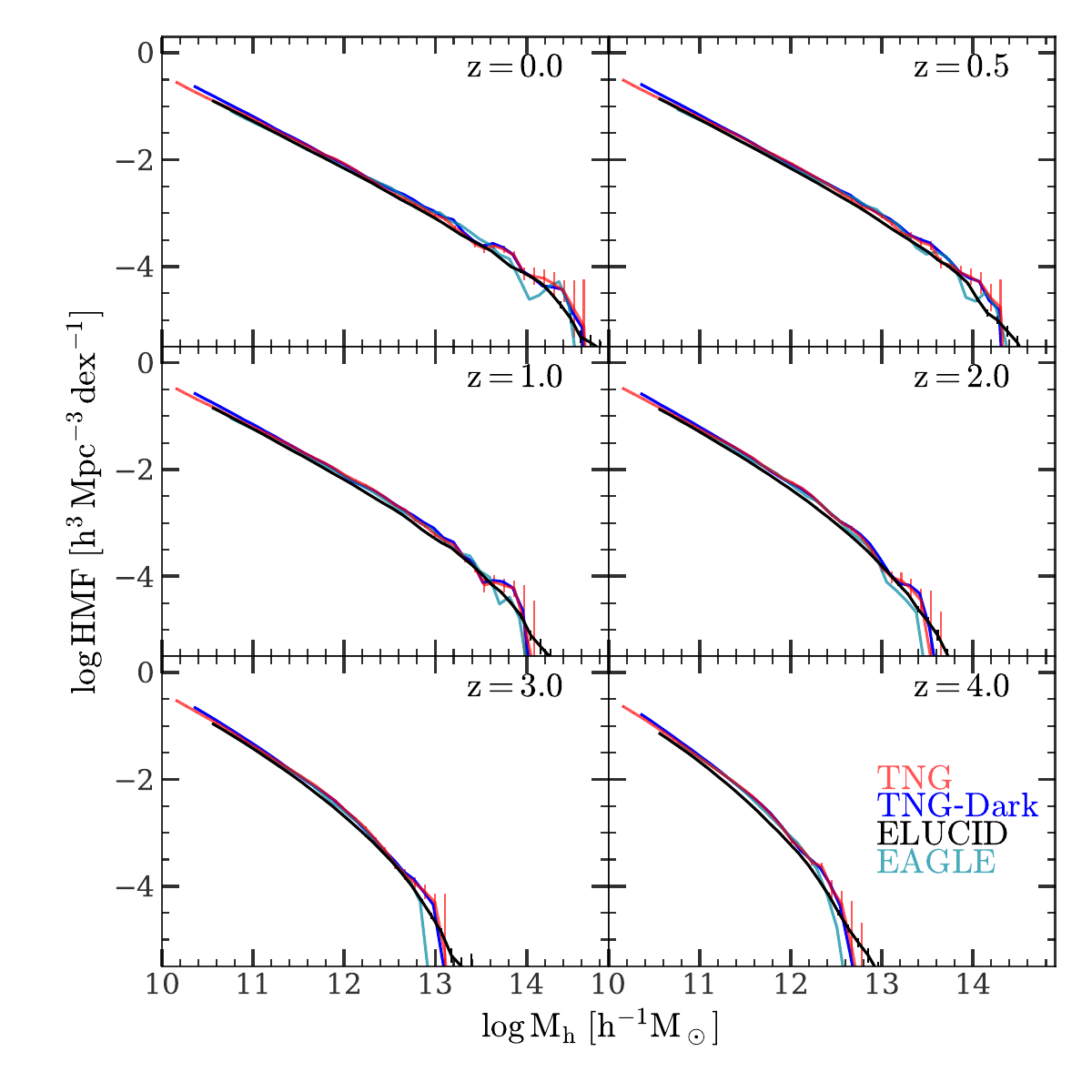}
	\caption{HMFs in TNG (\textbf{red}), TNG-Dark (\textbf{blue}), 
		ELUCID (\textbf{black}) and EAGLE (\textbf{green}) 
		simulations. Each panel 
		shows the HMFs at a given redshift indicated in the upper-right 
		corner of the panel. Error bars are computed by using 50 bootstrap 
		resamplings.}
	\label{fig:app-hmf}
\end{figure}

\begin{figure} \centering
	\includegraphics[width=\columnwidth]{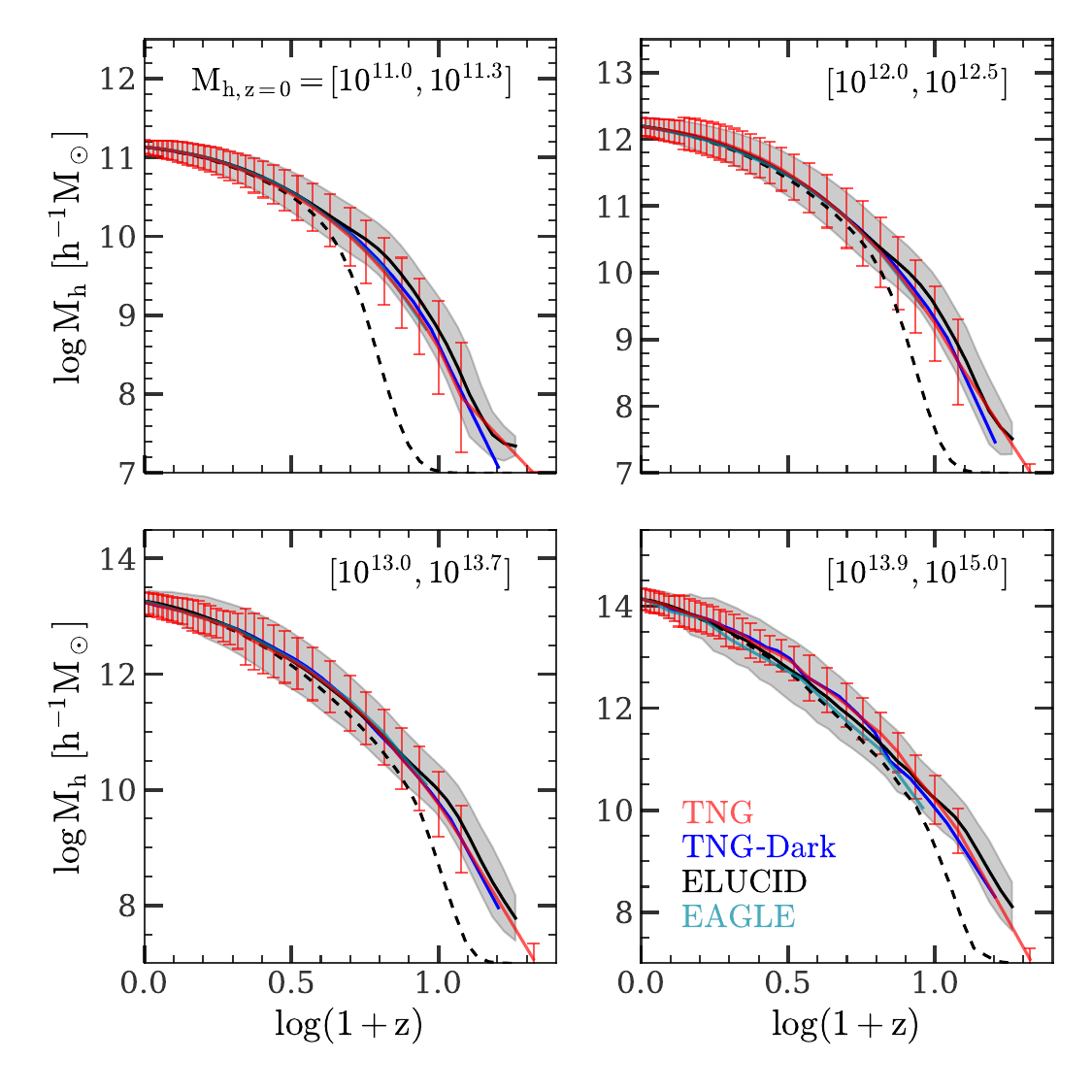}
	\caption{MAHs of $z=0$ central subhalos with different masses $\Mhalo / \msun$ as 
		indicated in each 
		panel. In each panel, the \textbf{red}, \textbf{blue} and \textbf{green} lines 
		are from the TNG, TNG-Dark and EAGLE simulations, 
		respectively. The \textbf{black solid} line is from the ELUCID simulation, 
		extended using TNG-Dark at high redshift. The \textbf{black dashed} 
		line is from the ELUCID simulation without any extension. 
		Error bars and 
		shades are standard deviations among halos in TNG and ELUCID 
		(with extension), respectively. }
	\label{fig:app-mah}
\end{figure}

\begin{figure} \centering
	\includegraphics[width=\columnwidth]{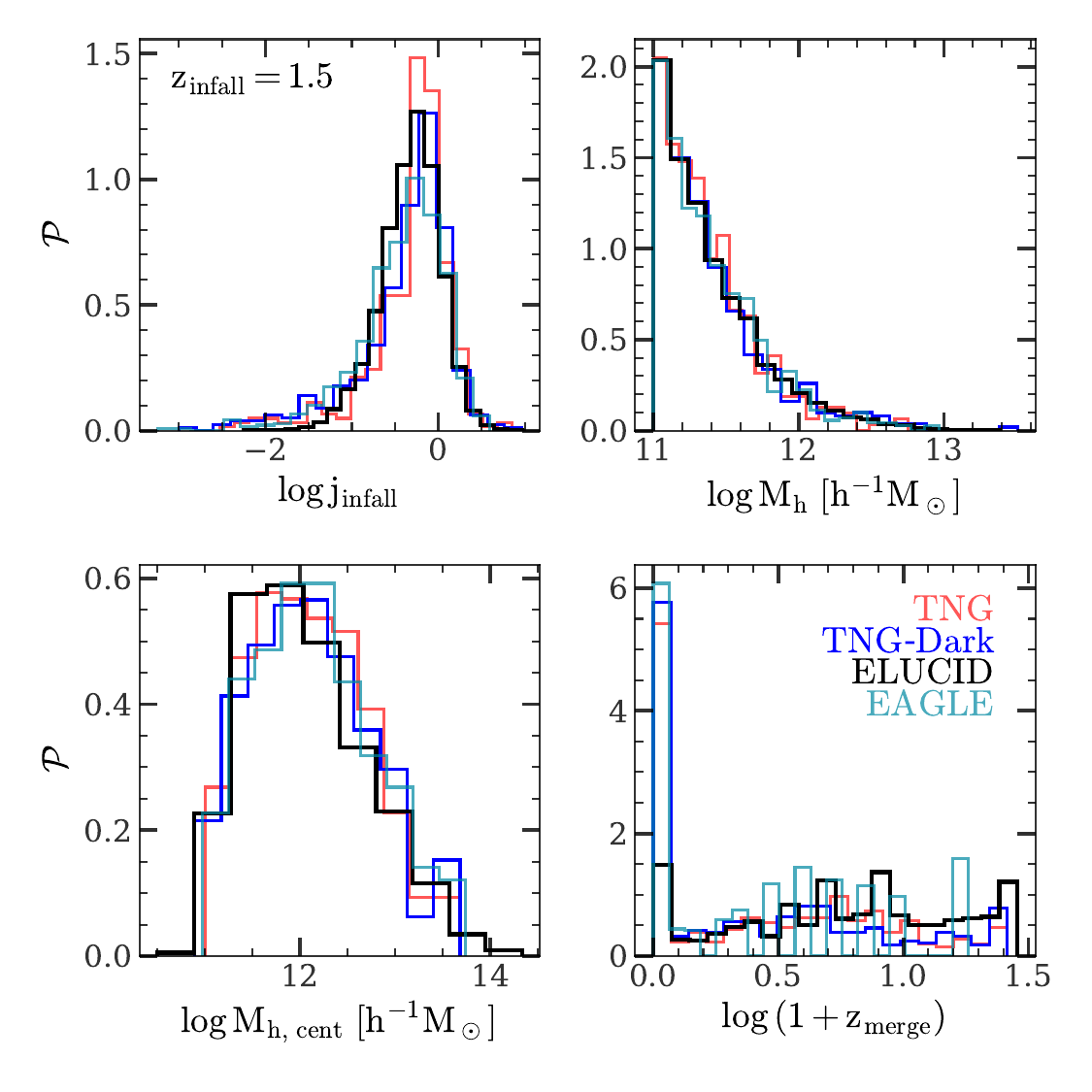}
	\caption{ Distributions of satellite subhalo properties. 
		The subhalo properties shown in the four panels are normalized 
		orbital angular momentum, satellite subhalo mass, subhalo mass of 
		the central subhalo that hosts the satellite, and the merger time. 
		The first three are computed at the infall-time.
		\textbf{Red}, \textbf{blue}, \textbf{black}, and \textbf{green} 
		histograms are from TNG, TNG-Dark,
		ELUCID and EAGLE simulations, respectively. 
		All subhalos with $z_{\rm infall}=1.5$ 
		are shown. }
	\label{fig:app-infall-dist}
\end{figure}

In this paper, we use dark matter halo/subhalos from TNG, TNG-Dark, ELUCID and 
EAGLE simulations as model input. These subhalos are mapped to galaxies using 
an interpretable deep model. In this appendix, we describe in detail the 
halo properties that are used in our analyses. One requirement of our model 
on the halo properties is that they must be stable, not sensitive to 
baryonic effects and numerical resolution (see \S\ref{ssec:overall-design}). 

Figure~\ref{fig:app-hmf} shows the halo mass functions (HMFs) at different 
redshifts in the three simulations. 
The HMFs of TNG, TNG-Dark and EAGLE almost overlap with each other
except at the high-mass end, where statistics are poor. This
indicates that the halo mass is a stable physical quantity that is not affected 
much by baryonic processes and not significantly different in 
different hydrodynamic simulations. 
The ELUCID HMFs are slightly lower than those 
of TNG, TNG-Dark and EAGLE. This is because ELUCID assumes a slightly 
different cosmology and has a lower mass resolution.

One advantage of first training an empirical model with small-volume, high-resolution 
hydrodynamic simulations and then applying it to large-volume DMO simulations 
is that more robust statistics can be drawn from the modeled galaxies in the DMO 
simulations. This can be seen in Figure~\ref{fig:app-hmf}, where the HMFS 
at the high-mass ends in TNG and EAGLE are both noisy, but 
the HMF of ELUCID remains stable and has a only small scatter 
even for halos with $M_{\rm h} > 10^{14} \msun$ at low $z$.

Since in our model galaxy properties are related not only to the 
current state of dark matter halos, but also to their mass assembly histories
(MAH) we, therefore, need to check the subhalo MAH given by different simulations.  
Figure~\ref{fig:app-mah} shows the MAH of $z=0$ subhalos with different masses.
The MAH in TNG, TNG-Dark and EAGLE have no obvious differences, indicating 
the robustness of this halo property used in our model.
Without any modification, the mass in the mean MAHs (black dashed lines) of 
ELUCID halos is significantly underestimated at high-$z$ 
when most halos are below the resolution limit of $6\times 10^9 \msun$.

As mentioned in \S\ref{ssec:samples}, we use halo merger trees 
given by TNG-Dark to complete the missing parts of the MAH of ELUCID
halos up to sufficiently high redshift. 
The extended MAHs, shown in Figure~\ref{fig:app-mah} 
as black solid lines, are comparable to the MAHs given by 
TNG, TNG-Dark and EAGLE. This extension largely eliminates the 
difference caused by numerical resolution, allowing the application 
of our model to the ELUCID simulation.

A number of properties for satellite subhalos are also used in our model, and here we 
check their reliability in different simulations. 
As an example, Figure~\ref{fig:app-infall-dist} shows the distributions of 
these properties for all subhalos with $\zinfall=1.5$. Only small differences 
can be seen in the distributions of $\log j_{\rm infall}$, $\Mhalo$, and 
$\log M_{\rm h,cent}$ between ELUCID and the other three simulations.
This can lead to some differences in the model predictions for satellite 
galaxies as seen in \S\ref{sec:results}. In contrast,
the merger time, $\zmerge$, obtained from ELUCID is quite 
different from those obtained from TNG, TNG-Dark and EAGLE.
Many subhalos in the high-resolution simulations can survive 
to $z=0$, while a large fraction of subhalos in ELUCID are
destroyed too early owing to the limited resolution. 
This affects our model by directly reducing the number of 
satellite subhalos. In \S\ref{ssec:model-of-satellite}, 
we solve this problem by using a model that predicts
the merger time from other, more robust halo properties.

\section{Training with and Application to the EAGLE Simulation}
\label{app:eagle-result}

\begin{figure*} \centering
	\includegraphics[width=16.5cm]{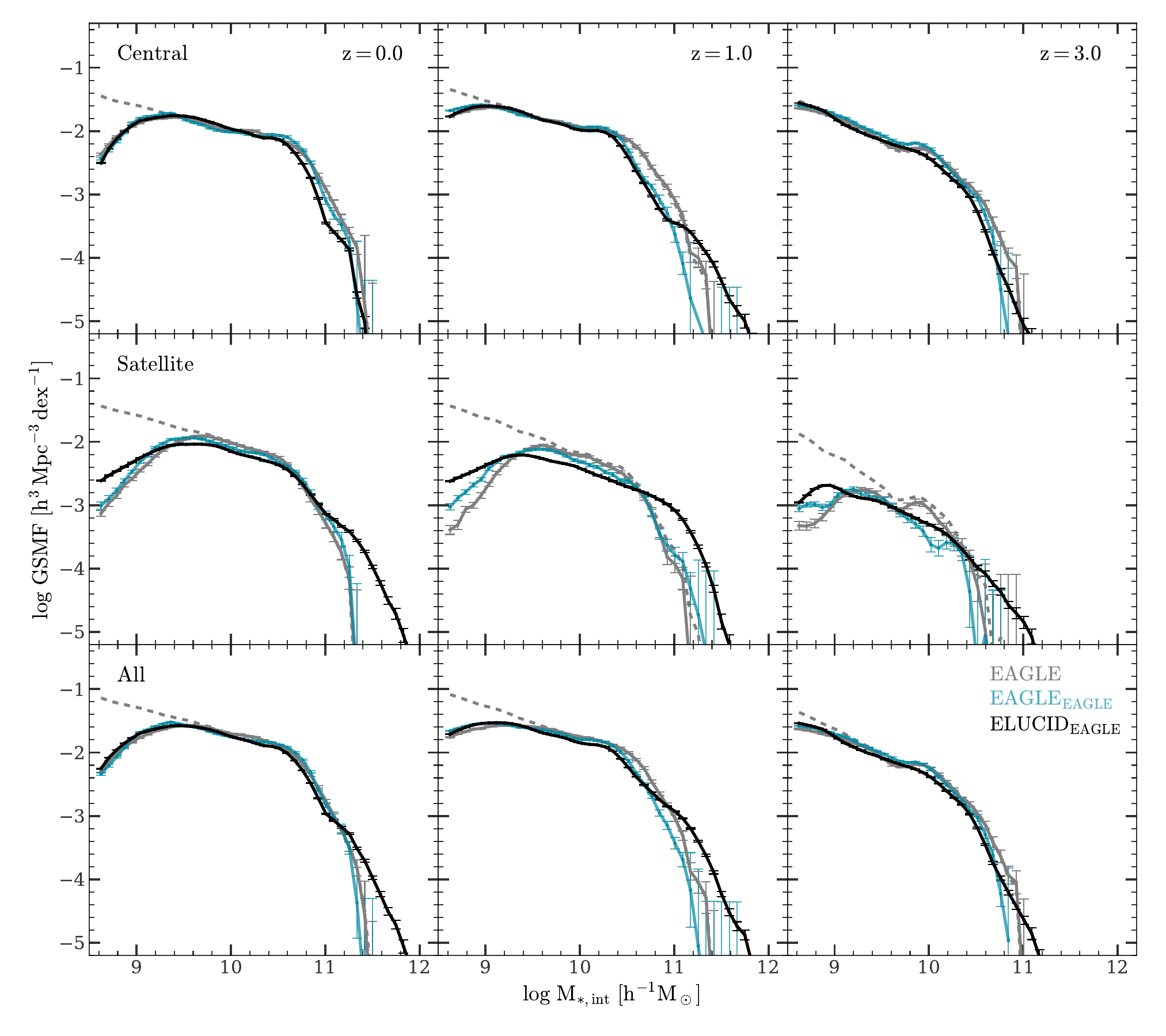}
	\caption{The GSMFs of modeled galaxies trained by EAGLE in comparison with 
	EAGLE simulated ones. 
	The \textbf{first}, \textbf{second} and \textbf{third} rows show 
	the GSMFs of central, satellite 
	and all (central+satellite) galaxies, respectively. 
	Different columns show the GSMFs at different redshifts. 
	\textbf{Solid gray} lines are 
	from the EAGLE simulation. The \textbf{green} and \textbf{black} lines are the results when 
	the model is applied to the EAGLE and the ELUCID simulations, respectively. 
	The \textbf{dashed gray} lines are GSMFs of all EAGLE galaxies including those in the 
	subhalos not selected in our samples. Error bars are computed 
	by using 50 bootstrap resamplings. }
	\label{fig:app-gsmf-eagle}
\end{figure*}

\begin{figure*} \centering
	\includegraphics[width=17cm]{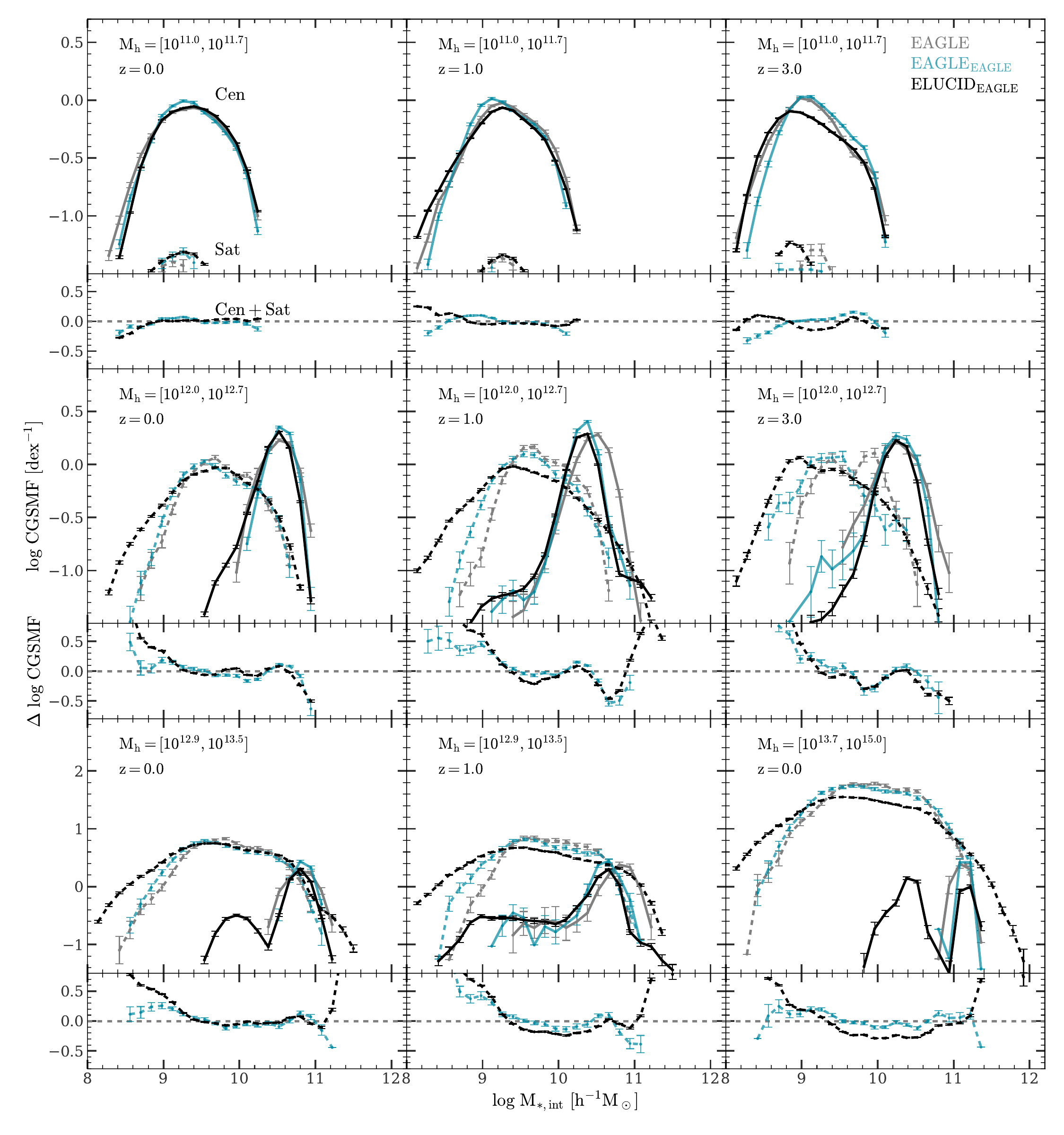}
	\caption{The CGSMFs of dark matter halos with different halo masses 
	$\Mhalo / (\msun)$ 
	and at different redshifts indicated in each panel. In each panel,
	the \textbf{gray} lines are 
	from the the EAGLE simulation. The \textbf{green} and 
	\textbf{black} lines are results when 
	the model is applied to EAGLE and ELUCID, respectively.  
	\textbf{Solid} and \textbf{dashed} 
	lines are for central and satellite galaxies, respectively.
	Error bars are computed by using 50 bootstrap resamplings. 
	{\revisestyle 
	Each small panel shows the model residuals of CGMSFs
	relative to the EAGLE simulation for all (central+satellite) galaxies.}
	}
	\label{fig:app-cgsmf-eagle}
\end{figure*}

In this appendix, we train \specialname[MAHGIC] using the EAGLE simulation, 
and then apply it to the subhalo merger trees obtained from 
both EAGLE and ELUCID. 
The model applied to EAGLE and ELUCID are referred to as 
$\rm EAGLE_{EAGLE}$ and 
$\rm ELUCID_{EAGLE}$, respectively. 

To achieve this, several changes are made to the modeling process. 
The merger tree branches in ELUCID are extended using the matched 
branches in EAGLE, instead of those in TNG-Dark (\S\ref{ssec:samples}).
The training data are taken from EAGLE and the application is also made to 
EAGLE, instead of TNG (\S\ref{sec:results}). 
All the other details, including the sample selection 
criteria, model pipeline 
and the separation redshifts, remain the same.

Figure~\ref{fig:app-gsmf-eagle} shows the GSMFs of EAGLE, $\rm EAGLE_{EAGLE}$
and $\rm ELUCID_{EAGLE}$ for central, satelite, and all galaxies at different 
redshifts from $0$ to $4$, respectively. As one can see,     
the GSMFs obtained from both $\rm EAGLE_{EAGLE}$ and $\rm ELUCID_{EAGLE}$
follow quite tightly those given by EAGLE. Overall, the performance of our 
model on EAGLE is very similar to its performance on TNG, indicating that 
the model is sufficiently flexible to accommodate different assumptions about
galaxy formation represented by the two simulations. 
We note that EAGLE has a smaller volume than TNG, so the cosmic variance is 
expected to be larger. This can be seen from Figure~\ref{fig:app-hmf}, 
where the high mass end of EAGLE HMF is underestimated at $z \geqslant 2$.
EAGLE also has a lower mass resolution and this can be seen from 
Figure~\ref{fig:app-mah}
where the MAH of halos stops at lower redshift. 
EAGLE has 29 output snapshots, less than the 100 snapshots in TNG,
so the training data are more limited. Taking these differences into 
consideration, the deviation of model results from EAGLE at 
the high stellar mass end of the GSMF can be explained. 
Figure~\ref{fig:app-cgsmf-eagle} shows the CGSMFs obtained from EAGLE, 
$\rm EAGLE_{EAGLE}$ and $\rm ELUCID_{EAGLE}$ for central and satellite galaxies.
Here we see again that our model performs well for halos of different mass, 
giving further support to its reliability and flexibility.

\bsp	
\label{lastpage}
\end{document}